\documentclass[acmsmall,nonacm]{acmart}

\AtBeginDocument{%
  }

\acmJournal{JACM}
\acmVolume{37}
\acmNumber{4}
\acmArticle{111}
\acmMonth{8}

\usepackage{forest}
\usepackage{tikz}
\usetikzlibrary{trees, matrix, patterns, positioning, arrows.meta}
\usepackage[utf8]{inputenc}
\usepackage{geometry}
\usepackage{setspace}
\usepackage{xcolor}
\usepackage{comment}
\usepackage{pgfplots}
\usepackage{filecontents}
\usepackage{tablefootnote}
\usepackage{amsmath}
\usepackage{graphicx}
\usepackage{url}
\usepackage[unicode]{hyperref}
\usepackage{subcaption}
\usepackage{amsfonts}

\usepackage[utf8]{inputenc}
\usepackage[T1]{fontenc}

\usepackage{color, soul}
\usepackage{mathrsfs}
\usepackage{cleveref}
\usepackage{multirow}
\usepackage{float}
\usepackage{wrapfig}
\usepackage{amsthm}
\usepackage{bm}
\usepackage{bbm}
\usepackage{algorithm}
\usepackage{algorithmic}
\usepackage{colortbl}
\usepackage{enumitem}
\usepackage{balance}
\usepackage{stfloats}
\usepackage{makecell}
\usepackage{dsfont}
\usepackage[T1]{fontenc}
\usepackage{booktabs}
\usepackage{nicefrac}
\usepackage{microtype}
\usepackage{amsmath}
\usepackage{amssymb}
\usepackage{bbding}
\usepackage{booktabs}
\usepackage{array}
\usepackage{makecell}
\usepackage{xcolor}
\usepackage{pifont}
\usepackage{natbib}
\usepackage{breakcites}
\usepackage{multirow}

\definecolor{connect-line}{RGB}{0,0,0}
\definecolor{middle-color}{RGB}{255,255,255}
\definecolor{leaf-color}{RGB}{255,255,255}
\definecolor{line-color}{RGB}{25,25,112}
\definecolor{extraction}{RGB}{240, 128, 128}
\definecolor{inversion}{RGB}{124, 205, 124}
\definecolor{others}{RGB}{135, 206, 235}
\definecolor{reliability}{RGB}{200, 200, 200}
\definecolor{generalizability}{RGB}{180, 180, 180}

\definecolor{attack}{RGB}{255,200,200}
\definecolor{defense}{RGB}{200,255,200}
\definecolor{environment}{RGB}{200,200,255}

\definecolor{lightgray}{HTML}{e0e0e0}

\tikzset{
  forked edges/.style={},
  grey/.style={fill=gray!30},
  extraction-middle/.style={draw=extraction, fill=middle-color!40, text opacity=1, align=center, fill opacity=.5, text=black, font=\scriptsize, inner sep=3pt},
  extraction-leaf/.style={draw=extraction, fill=leaf-color!40, text opacity=1, align=left, fill opacity=.5, text=black, font=\scriptsize, inner sep=3pt},
  inversion-middle/.style={draw=inversion, fill=middle-color!40, text opacity=1, align=center, fill opacity=.5, text=black, font=\scriptsize, inner sep=3pt},
  inversion-leaf/.style={draw=inversion, fill=leaf-color!40, text opacity=1, align=left, fill opacity=.5, text=black, font=\scriptsize, inner sep=3pt},
  others-middle/.style={draw=others, fill=middle-color!40, text opacity=1, align=center, fill opacity=.5, text=black, font=\scriptsize, inner sep=3pt},
  others-leaf/.style={draw=others, fill=leaf-color!40, text opacity=1, align=left, fill opacity=.5, text=black, font=\scriptsize, inner sep=3pt},
  reliability-middle/.style={draw=reliability, fill=middle-color!40, text opacity=1, align=center, fill opacity=.5, text=black, font=\scriptsize, inner sep=3pt},
  reliability-leaf/.style={draw=reliability, fill=leaf-color!40, text opacity=1, align=left, fill opacity=.5, text=black, font=\scriptsize, inner sep=3pt},
  generalizability-middle/.style={draw=generalizability, fill=middle-color!40, text opacity=1, align=center, fill opacity=.5, text=black, font=\scriptsize, inner sep=3pt},
  generalizability-leaf/.style={draw=generalizability, fill=leaf-color!40, text opacity=1, align=left, fill opacity=.5, text=black, font=\scriptsize, inner sep=3pt}
}

\forestset{
  default preamble={
    for tree={
      align=center,
      parent anchor=south,
      child anchor=north,
    }
  },
}

\begin{document}


\title{A Systematic Survey of Model Extraction Attacks and Defenses: State-of-the-Art and Perspectives}

\author{Kaixiang Zhao}
\email{kzhao5@nd.edu}
\orcid{0009-0005-8174-0581}
\affiliation{%
  \institution{University of Notre Dame}
  \city{South Bend}
  \state{Indiana}
  \country{USA}
}

\author{Lincan Li}
\email{ll24bb@fsu.edu}
\orcid{0000-0003-3797-4055}
\affiliation{%
  \institution{Florida State University}
  \city{Tallahassee}
  \state{Florida}
  \country{USA}
}

\author{Kaize Ding}
\email{kaize.ding@northwestern.edu}
\orcid{0000-0001-6684-6752}
\affiliation{%
  \institution{Northwestern University}
  \city{Evanston}
  \state{Illinois}
  \country{USA}
}

\author{Neil Zhenqiang Gong}
\email{neil.gong@duke.edu}
\orcid{0000-0002-9900-9309}
\affiliation{%
  \institution{Duke University}
  \city{Durham}
  \state{North Carolina}
  \country{USA}
}

\author{Yue Zhao}
\email{yzhao010@usc.edu}
\orcid{0000-0003-3401-4921}
\affiliation{%
  \institution{University of Southern California}
  \city{Los Angeles}
  \state{California}
  \country{USA}
}

\author{Yushun Dong}
\authornote{Corresponding author}
\email{yd24f@fsu.edu}
\orcid{0000-0001-7504-6159}
\authornotemark[0]
\affiliation{%
  \institution{Florida State University}
  \city{Tallahassee}
  \state{Florida}
  \country{USA}
}

\renewcommand{\shortauthors}{Zhao et al.}

\newcommand\neil[1]{\textcolor{blue}{[NG: #1]}}

\begin{abstract}

Machine learning (ML) models have significantly grown in complexity and utility, driving advances across multiple domains. However, substantial computational resources and specialized expertise have historically restricted their wide adoption. Machine-Learning-as-a-Service (MLaaS) platforms have addressed these barriers by providing scalable, convenient, and affordable access to sophisticated ML models through user-friendly APIs. While this accessibility promotes widespread use of advanced ML capabilities, it also introduces vulnerabilities exploited through Model Extraction Attacks (MEAs). Recent studies have demonstrated that adversaries can systematically replicate a target model's functionality by interacting with publicly exposed interfaces, posing threats to intellectual property, privacy, and system security. In this paper, we offer a comprehensive survey of MEAs and corresponding defense strategies. We propose a novel taxonomy that classifies MEAs according to attack mechanisms, defense approaches, and computing environments. Our analysis covers various attack techniques, evaluates their effectiveness, and highlights challenges faced by existing defenses, particularly the critical trade-off between preserving model utility and ensuring security. We further assess MEAs within different computing paradigms and discuss their technical, ethical, legal, and societal implications, along with promising directions for future research. This systematic survey aims to serve as a valuable reference for researchers, practitioners, and policymakers engaged in AI security and privacy. Additionally, we maintain an online repository continuously updated with related literature at https://github.com/kzhao5/ModelExtractionPapers.
\end{abstract}



\maketitle

\section{Introduction}

The emergence of Machine-Learning-as-a-Service (MLaaS) has significantly improved the accessibility of AI technologies by lowering the barriers traditionally associated with adopting and deploying complex machine learning models \cite{yao2017complexity}. Initially, deploying machine learning required considerable investments in computational infrastructure, extensive data storage, and specialized expertise to develop, train, and maintain models. These requirements inherently limited the adoption of advanced AI capabilities to organizations with substantial resources. MLaaS platforms address these limitations by abstracting the complexity of model development and deployment through user-friendly, standardized APIs. Consequently, organizations and developers, irrespective of their size or technical proficiency, can integrate sophisticated AI models into applications without direct involvement in model training, hardware provisioning, or scalability management. This ease of access has facilitated widespread adoption across various sectors, including healthcare \cite{khosravi2022new,javaid2022significance} and finance \cite{javaid2022significance, abdelaziz2018machine, ribeiro2015mlaas}, effectively democratizing AI for smaller and less technically-equipped entities. However, the publicly accessible interfaces of MLaaS, essential for their integration, inadvertently introduce new security vulnerabilities \cite{tramer2016stealing, papernot2017practical}. Specifically, adversaries can exploit the query-response mechanisms fundamental to MLaaS to systematically probe and infer critical information about model architectures, parameters, and decision boundaries. These model extraction attacks (MEAs) pose significant threats, potentially compromising the intellectual property of private models and enabling unauthorized replication or misuse \cite{shokri2017membership, orekondy2019knockoff, krishna2019thieves}.

The exposed interfaces of MLaaS platforms, although essential for their operation, inherently introduce a significant security vulnerability: adversaries can perform MEAs by exploiting the fundamental query-response mechanism provided by these interfaces. Specifically, attackers repeatedly interact with the target model through carefully crafted queries to systematically probe its decision boundaries and response behaviors \cite{tramer2016stealing, jagielski2020high, papernot2017practical}. By analyzing the model’s responses, attackers accumulate knowledge about its architecture, parameters, and decision-making processes, ultimately constructing a substitute model that closely replicates the original model's functionality \cite{orekondy2019knockoff, krishna2019thieves, shokri2017membership}. This extraction process becomes increasingly efficient through iterative refinement, where insights from previous queries are leveraged to design more informative subsequent queries, thus reducing the total query count \cite{jagielski2020high, shokri2017membership}. Recent developments have revealed the severity and breadth of MEAs threats across multiple domains. In language models, research has demonstrated the possibility of replicating GPT-3.5-like performance with minimal resources \cite{anand2023gpt4all}, and subsequent studies showed that extraction could be executed affordably through optimized prompts \cite{carlini2024stealing}, fostering malicious LLM variants such as WormGPT and FraudGPT for illegal activities \cite{Falade_2023}. Similarly concerning trends have emerged in autonomous driving, where core components like object detection and path planning models can be reconstructed via black-box access \cite{orekondy2019knockoff, Hu2020DeepSniffer}, facilitating adversarial modifications to traffic signs \cite{wu2019adversarial, breier2022sniff}. Edge computing and federated learning systems are also vulnerable, as attackers exploit side-channel vulnerabilities through electromagnetic \cite{Batina2019reverse} or cache timing analysis \cite{Yan2020cache}. These impacts span various architectures, including vision-language models \cite{wu2022model}, reinforcement learning \cite{chen2021stealing}, and diffusion models \cite{duan2023are}, with extracted models enabling sensitive data inference \cite{carlini2021extracting, li2022data, zhang2021graphmi, huang2022are} or facilitating targeted attacks \cite{yan2023explanation, oksuz2023autolycus}. The financial sector has shown particular vulnerability, evidenced by attackers reconstructing prediction and risk assessment models via systematic API queries \cite{karmakar2023marich, tramer2017ensemble}. A striking real-world example of these vulnerabilities involved a \$35 million voice fraud case in the UAE, highlighting how extracted models combined with biometric data can enable sophisticated fraud \cite{unknown}. Collectively, these vulnerabilities underscore that MEAs represent a systemic threat to the MLaaS ecosystem rather than isolated incidents \cite{shokri2017membership, krishna2019thieves}.

Despite substantial progress, several critical challenges remain unresolved in the field of model extraction attacks and defenses. Firstly, there is a significant gap in the systematic understanding of extraction attack methodologies. A comprehensive understanding of these attacks across various model architectures, data modalities, and computing paradigms is crucial for developing more effective defense strategies. Current theoretical frameworks are insufficient for quantifying the fundamental limits of model extractability under different constraints. Secondly, the defense landscape itself is fragmented and complex, with numerous approaches available—ranging from perturbation and watermarking to anomaly detection and differential privacy—each having unique advantages and limitations. This complexity makes it challenging for practitioners to select appropriate defenses tailored to specific contexts and threat scenarios. Additionally, balancing effective model protection with minimal performance degradation is particularly difficult in high-stakes applications requiring stringent security and accuracy constraints. The rapid evolution of sophisticated attack methodologies leveraging advanced optimization techniques and generative models further necessitates continuous innovation in defense strategies. Moreover, the ethical, legal, and societal implications of model extraction remain underexplored, especially concerning intellectual property rights and the malicious utilization of extracted models. Addressing these challenges is essential to bridge the existing gap between advancements in extraction attacks and the development of robust, adaptable defense solutions.

In view of this, this paper aims to provide a comprehensive and up-to-date overview of model extraction attacks and defenses. Our main goal is to establish a unified framework that comprehensively analyzes all aspects of MEAs, including attack techniques, defense strategies, application scenarios, and potential impacts. Following this general introduction, we first present essential background knowledge and fundamental concepts in MEAs. We then propose a novel and comprehensive taxonomy to systematically classify MEAs based on attack mechanisms, defense strategies, and computing environments. Subsequently, we provide a detailed, systematic review of current techniques for both attacks and defenses, highlighting their methodologies, advantages, disadvantages, and the conditions under which they are most effective. Further, we discuss MEAs and corresponding defense strategies across different computing paradigms, including centralized, cloud computing, edge computing, and federated learning, illustrating their specific challenges and potential solutions. Additionally, we assess the implications of these attacks comprehensively from technical, ethical, legal, and societal perspectives. Our analysis identifies key research challenges and outlines promising future directions, aiming to bridge current gaps and advance the development of robust defense mechanisms. Finally, we have created an online repository to continuously update and aggregate related papers and resources in the model extraction field, assisting researchers in staying informed about the latest developments. Our research not only covers technical challenges but also explores ethical, legal, and social issues, providing guidance for responsible AI development.

To summarize, the contributions of this paper are four-fold:
\begin{itemize}
\item \textbf{A Comprehensive Taxonomy of Model Extraction.} We present a novel and comprehensive taxonomy that systematically classifies MEAs based on different attack mechanisms, defense strategies, and computing environments.
\item \textbf{Systematic Review of Current Techniques.} We systematically review current techniques for model extraction attacks and defenses, providing detailed insights into their methodologies, advantages, disadvantages, and conditions under which each technique is most effective.
\item \textbf{Multi-dimensional Assessment and Future Perspectives.} We assess the impact of MEAs from technical, ethical, legal, and societal dimensions and outline critical research challenges and promising future research directions.
\item \textbf{Online Updating Resource.} We maintain an open-source repository that continuously collects related papers in the field of model extraction, providing resources such as paper links, code repositories, benchmarks, and performance comparisons.
\end{itemize}

\noindent \textbf{Difference from Other Related Surveys.} Many researchers have extensively explored the field of MEAs and related security challenges. General privacy studies in machine learning \cite{de2021critical, rigaki2023survey, he2022towards, liu2021machine, liu2020privacy, tanuwidjaja2020privacy} primarily discuss various privacy preservation techniques but do not specifically focus on MEAs and their distinctive challenges. Additionally, domain-specific research \cite{ju2024survey, guan2024graph, wang2024safety, zhang2023survey, wang2024unique, dai2023comprehensive, yao2024survey, zhao2025survey2} has examined privacy and security issues within particular contexts, such as graph and text data analysis, but these studies offer limited insights into the broader implications of model extraction across diverse model architectures and scenarios. Other works \cite{nayan2024sok, lyu2022privacy, qayyum2020securing, zhang2023survey, zhao2025survey} have investigated MEAs in specific computing environments, often presenting isolated analyses without adequately addressing interactions among different computing paradigms. Recent surveys, such as those by \cite{gencc2023taxonomic} and \cite{oliynyk2023know}, mainly catalog existing attack methodologies but do not sufficiently provide deeper analytical insights, such as comparative effectiveness, limitations, or comprehensive frameworks integrating both attacks and defenses. As illustrated in Table~\ref{tab:survey_comparison}, our survey differs from these existing works by offering the first unified and comprehensive framework that systematically addresses these limitations. Specifically, we provide a detailed and integrated examination of MEAs, clearly identifying their unique challenges, systematically categorizing them through a novel taxonomy, and thoroughly reviewing attack-defense dynamics across multiple computing environments. This approach offers researchers and practitioners targeted insights and a thorough understanding of the model extraction landscape, thereby aiding in the development of robust defenses and secure machine learning systems.

\begin{table}[!t]
\centering
\small
\caption{Comparison of survey works on model extraction attacks and defenses.}
\setlength{\tabcolsep}{2pt}
\renewcommand{\arraystretch}{1.2}
\rowcolors{2}{lightgray}{}
\resizebox{\textwidth}{!}{
\begin{tabular}{l|c|c|c|c|c|c|c|c|c}
\hline
 & \makecell[c]{Literature Review \\ from General \\ Machine Learning} 
 & \makecell[c]{Proposed\\Taxonomy on \\Attack \& Defense} 
 & \makecell[c]{Analysis across \\ Computing \\ Environments} 
 & \makecell[c]{Multiple Data \\ Modalities \\ Assessment} 
 & \makecell[c]{Evaluation \\ Protocols} 
 & \makecell[c]{Real-world \\ Application \\ Scenarios} 
 & \makecell[c]{Analyzing \\ Limitations \& \\ Challenges} 
 & \makecell[c]{Future\\Research \\ Direction} 
 & \makecell[c]{Online\\Updating \\ Resource} \\
\hline
\cite{de2021critical} & \textcolor{red}{\ding{55}} & \textcolor{red}{\ding{55}} & \textcolor{green}{\ding{51}} & \textcolor{red}{\ding{55}} & \textcolor{green}{\ding{51}} & \textcolor{green}{\ding{51}} & \textcolor{green}{\ding{51}} & \textcolor{green}{\ding{51}} & \textcolor{green}{\ding{51}} \\
\cite{rigaki2023survey} & \textcolor{red}{\ding{55}} & \textcolor{red}{\ding{55}} & \textcolor{red}{\ding{55}} & \textcolor{red}{\ding{55}} & \textcolor{green}{\ding{51}} & \textcolor{green}{\ding{51}} & \textcolor{green}{\ding{51}} & \textcolor{green}{\ding{51}} & \textcolor{red}{\ding{55}} \\
\cite{he2022towards} & \textcolor{red}{\ding{55}} & \textcolor{red}{\ding{55}} & \textcolor{green}{\ding{51}} & \textcolor{red}{\ding{55}} & \textcolor{green}{\ding{51}} & \textcolor{green}{\ding{51}} & \textcolor{green}{\ding{51}} & \textcolor{green}{\ding{51}} & \textcolor{red}{\ding{55}} \\
\cite{liu2021machine} & \textcolor{red}{\ding{55}} & \textcolor{red}{\ding{55}} & \textcolor{red}{\ding{55}} & \textcolor{red}{\ding{55}} & \textcolor{green}{\ding{51}} & \textcolor{green}{\ding{51}} & \textcolor{green}{\ding{51}} & \textcolor{green}{\ding{51}} & \textcolor{red}{\ding{55}} \\
\cite{liu2020privacy} & \textcolor{red}{\ding{55}} & \textcolor{red}{\ding{55}} & \textcolor{red}{\ding{55}} & \textcolor{red}{\ding{55}} & \textcolor{green}{\ding{51}} & \textcolor{green}{\ding{51}} & \textcolor{green}{\ding{51}} & \textcolor{green}{\ding{51}} & \textcolor{red}{\ding{55}} \\
\cite{tanuwidjaja2020privacy} & \textcolor{red}{\ding{55}} & \textcolor{red}{\ding{55}} & \textcolor{red}{\ding{55}} & \textcolor{red}{\ding{55}} & \textcolor{green}{\ding{51}} & \textcolor{green}{\ding{51}} & \textcolor{green}{\ding{51}} & \textcolor{green}{\ding{51}} & \textcolor{red}{\ding{55}} \\
\cite{ju2024survey} & \textcolor{red}{\ding{55}} & \textcolor{red}{\ding{55}} & \textcolor{red}{\ding{55}} & \textcolor{red}{\ding{55}} & \textcolor{green}{\ding{51}} & \textcolor{green}{\ding{51}} & \textcolor{green}{\ding{51}} & \textcolor{green}{\ding{51}} & \textcolor{red}{\ding{55}} \\
\cite{guan2024graph} & \textcolor{red}{\ding{55}} & \textcolor{red}{\ding{55}} & \textcolor{red}{\ding{55}} & \textcolor{red}{\ding{55}} & \textcolor{green}{\ding{51}} & \textcolor{green}{\ding{51}} & \textcolor{green}{\ding{51}} & \textcolor{green}{\ding{51}} & \textcolor{red}{\ding{55}} \\
\cite{wang2024safety} & \textcolor{red}{\ding{55}} & \textcolor{red}{\ding{55}} & \textcolor{red}{\ding{55}} & \textcolor{red}{\ding{55}} & \textcolor{green}{\ding{51}} & \textcolor{green}{\ding{51}} & \textcolor{green}{\ding{51}} & \textcolor{green}{\ding{51}} & \textcolor{red}{\ding{55}} \\
\cite{zhang2023survey} & \textcolor{red}{\ding{55}} & \textcolor{red}{\ding{55}} & \textcolor{red}{\ding{55}} & \textcolor{red}{\ding{55}} & \textcolor{green}{\ding{51}} & \textcolor{green}{\ding{51}} & \textcolor{green}{\ding{51}} & \textcolor{green}{\ding{51}} & \textcolor{red}{\ding{55}} \\
\hline
\makecell[c]{{}\\ \textbf{Ours}}& \textcolor{green}{\ding{51}} & \textcolor{green}{\ding{51}} & \textcolor{green}{\ding{51}} & \textcolor{green}{\ding{51}} & \textcolor{green}{\ding{51}} & \textcolor{green}{\ding{51}} & \textcolor{green}{\ding{51}} & \textcolor{green}{\ding{51}} & \textcolor{green}{\ding{51}} \\
\hline
\end{tabular}}
\label{tab:survey_comparison}
\end{table}

\noindent\textbf{Intended Audiences.} 

Our comprehensive survey serves multiple intended audiences. Researchers in the field will find detailed insights into recent advances, techniques, and challenges related to MEAs, facilitating further academic inquiry and innovation. AI and machine learning engineers will benefit from understanding practical strategies to protect their models against extraction attacks, enabling the effective implementation of robust defenses in real-world applications. Additionally, policymakers and legal professionals concerned with AI security and privacy protection can leverage this survey to inform the development of regulations, policies, and frameworks aimed at safeguarding against potential risks associated with model extraction threats.

\noindent\textbf{Survey Structure.} In Section \ref{background}, we introduce relevant background knowledge and basic concepts to provide readers with the necessary theoretical foundation.
In Section \ref{taxonomy}, we propose a novel taxonomy of MEAs, and in Section \ref{MEA} we explore various attack techniques in detail. In Section \ref{MED}, we discuss the corresponding defense strategies.
In Section \ref{environment}, we also discuss MEAs studies in different computing environments.
In Section \ref{evaluation}, we explore the basic and advanced evaluation metrics for attacks and defenses. 
In Section \ref{app}, we analyze attack and defense cases in real-world application scenarios to provide readers with practical references.
In Section \ref{future}, we discuss future research directions and challenges, including interdisciplinary topics such as legal and ethical issues. 
Finally, Section \ref{conclusion} presents the conclusion of this survey.

\noindent

\section{Background and Preliminaries} \label{background}
In this section, we discuss the essential background knowledge and preliminaries. We give the notation and terminology first. Then, we provide fundamentals of MLaaS. Next, we provide model extraction attack basics including problem settings, threat model. Finally, We provide detailed introduction to the defense of MEAs.

\subsection{Notation and Terminology}
\textbf{Notations.} $\mathcal{S}$ denotes the dataset, typically represented as $\{(x_i, y_i)\}_{i=1}^n$, where $\mathcal{X}$ is the input space, $\mathcal{Y}$ is the output space, and $n$ is the number of samples. The hypothesis space is denoted by $\mathcal{H}$, with $f_{\theta}(x)$ representing a hypothesis function. We use $\ell$ for the loss function, $L_{\mathcal{S}}(\theta)$ for the empirical risk, and $L_{\mathcal{D}}(\theta)$ for the expected risk. For matrices and vectors, we use $\bm{A}$ and $\bm{x}$ respectively, with $\bm{A}^\top$ denoting transpose. In the context of MEAs, $\mathcal{M}$ represents the target model and $\mathcal{M}'$ the substitute model. For a comprehensive list of notations used throughout this survey, including those specific to different machine learning paradigms and computing environments, please refer to Table \ref{tab:notation}.

\noindent \textbf{Terminology.} To ensure clarity and consistency throughout this paper, we use specific terminology to describe various aspects of MEAs. For a comprehensive list of terms and their definitions, please refer to Table \ref{tab:terminology}, which provides a detailed overview of the terminology used in the context of model extraction attacks.

\begin{table}[htbp]
\centering
\small
\setlength{\tabcolsep}{3pt}
\rowcolors{2}{lightgray}{}
\begin{tabular}{p{0.1\textwidth}|p{0.37\textwidth}||p{0.1\textwidth}|p{0.37\textwidth}}
\hline
\textbf{Notation} & \textbf{Definition or Description} & \textbf{Notation} & \textbf{Definition or Description} \\
\hline
$|\cdot|$ & Cardinality operator for any set & $\mathcal{M}$ & Target model in extraction attacks \\
$E[\cdot]$ & Expectation operator & $\mathcal{M}'$ & Substitute model in extraction attacks \\
$\mathcal{S}$ & Dataset, typically $\{(x_i, y_i)\}_{i=1}^n$ & $Q$ & Query set for model extraction \\
$\mathcal{X}$ & Input space or feature space & $B$ & Query budget for extraction attacks \\
$\mathcal{Y}$ & Output space or label space & $\mathcal{A}$ & Extraction attack algorithm \\
$n$ & Number of samples in the dataset & $\mathcal{D}$ & Defense mechanism against extraction \\
$\mathcal{H}$ & Hypothesis space, set of possible models & $\mathcal{L}$ & Language model (e.g., BERT, GPT) \\
$f_{\theta}(x)$ & Hypothesis function with parameters $\theta$ & $\mathcal{T}$ & Task-specific fine-tuning of LLM \\
$f^*$ & Target function to be learned & $I$ & Image in computer vision tasks \\
$\ell$ & Loss function measuring prediction error & $\mathbf{F}$ & Feature map in CNNs \\
$L_{\mathcal{S}}(\theta)$ & Empirical risk on dataset $\mathcal{S}$ & $\mathcal{C}$ & Set of clients in federated learning \\
$L_{\mathcal{D}}(\theta)$ & Expected risk over distribution $\mathcal{D}$ & $K$ & Number of clients in federated learning \\
$\sigma(x)$ & Activation function in neural networks & $\mathcal{G}$ & Graph, typically $(\mathcal{V}, \mathcal{E})$ \\
$d$ & Dimension of features or embeddings & $\mathcal{V}$ & Set of nodes in a graph \\
$c$ & Number of classes in classification tasks & $\mathcal{E}$ & Set of edges in a graph \\
$\mathbf{w}$ & Word in text processing & $\mathbf{A}$ & General matrix \\
$\mathbf{s}$ & Sentence or sequence in text processing & $\mathbf{x}$ & General vector \\
$\mathcal{P}$ & Prompt in LLM interactions & $\mathbf{A}^\top$ & Matrix transpose \\
$\mathcal{R}$ & Response from LLM & $\theta$ & Model parameters \\
$\alpha$ & Learning rate in optimization & $w_k$ & Weight of client $k$ in federated learning \\
$\epsilon$ & Small perturbation in adversarial attacks & $\eta$ & Extraction attack success rate \\
$\lambda$ & Regularization parameter & $\delta$ & Confidence level in statistical tests \\
$\mathbf{W}$ & Weight matrix in neural networks & $\mathbf{b}$ & Bias vector in neural networks \\
$h$ & Hidden layer in neural networks & $o$ & Output layer in neural networks \\
$\nabla$ & Gradient operator & $\mathcal{O}$ & Big O notation for complexity \\
$L_k$ & Local loss function for client $k$ & & \\
\hline
\end{tabular}
\caption{Notations and the corresponding descriptions.}
\label{tab:notation}
\end{table}

\begin{table}[htbp]
\centering
\small
\rowcolors{2}{lightgray}{}
\begin{tabular}{p{0.25\textwidth}p{0.7\textwidth}}
\hline
\textbf{Term} & \textbf{Definition} \\
\hline
Model Extraction Attack & Attack aiming to replicate functionality of a target machine learning model \\
\hline
Target Model & Original model targeted by the extraction attack \\
\hline
Substitute Model & Model trained by adversary to mimic the target model \\
\hline
Query-based Attack & Attack method obtaining information by querying the target model \\
\hline
MLaaS & Cloud services providing machine learning functionalities via APIs \\
\hline
API & Interface allowing interaction with the target model \\
\hline
Adversary & Entity performing the model extraction attack \\
\hline
Query Generation & Process of creating input data to probe the target model \\
\hline
substitute Dataset & Dataset for generating queries and training the substitute model \\
\hline
Transfer Dataset & Dataset used to transfer knowledge from target to substitute model \\
\hline
Black-box Attack & Attack without knowledge of target model's internal structure \\
\hline
White-box Attack & Attack with full knowledge of target model's internal structure \\
\hline
Fidelity & Measure of how closely substitute model matches target model \\
\hline
Query Efficiency & Ability to gain maximum information with minimum queries \\
\hline
\end{tabular}
\caption{Terminology for Model Extraction Attacks}
\label{tab:terminology}
\end{table}

\subsection{Basics of Machine-Learning-as-a-Service}
\noindent\textbf{Definition.} Machine-Learning-as-a-Service (MLaaS)~\cite{ribeiro2015mlaas} refers to cloud-based computing platforms that offer machine learning tools and capabilities. These machine learning services, provided by cloud computing providers like Amazon, Google, Microsoft Azure, allow users to train or finetune models, evaluate them, or utilize pre-trained models, typically via a pay-per-query model. The key benefit of MLaaS is to provide computing resources and machine learning capabilities without requiring users to manage the underlying infrastructure or have deep expertise in machine learning algorithms. Users can access machine learning resources efficiently from anywhere, at any time, focusing mainly on the data and problem at hand rather than implementation details~\cite{tramer2016stealing}.

\noindent\textbf{Service Architecture.} MLaaS platforms usually provide functionalities such as data pre-processing, model training, evaluation and validation, etc. These functionalities usually provide through APIs that allow only input-output interactions, without revealing the model architecture or parameters. These platforms provide two primary types of APIs: training APIs for custom model development and prediction APIs for generating predictions using pre-trained or custom models.

\noindent\textbf{Security Implications.} New security considerations related to MEAs introduced by the special features of MLaaS paradigm. The black-box nature of most MLaaS offerings provides some inherent protection against model extraction\cite{kesarwani2018model}. Service providers can implement query monitoring systems to detect malicious queries, use output perturbation techniques, and offer limited access control\cite{hou2019ml,ribeiro2015mlaas}. However, MLaaS also presents challenges for model protection, such as an increased attack vulnerability due to wide accessibility, resource constraints for implementing comprehensive defenses at scale, and trade-offs between model protection and service quality.

\subsection{Basics of Model Extraction Attacks}

\begin{figure}[htbp]
    \centering
    \hspace*{-0.03\textwidth}
    \includegraphics[width=1.05\textwidth]{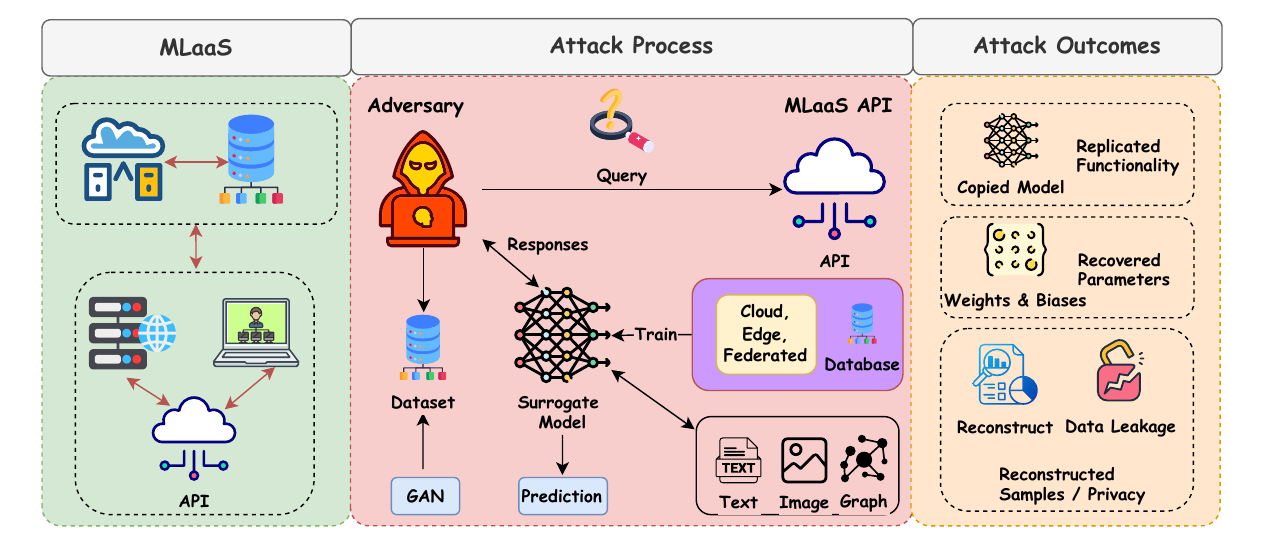}
    \caption{Model Extraction Attack Overview.}
    \label{fig:mea_overview}
\end{figure}

MEAs aim to infer a target model's properties, typically through black-box query access. The inferred information may include the model's architecture, (hyper)parameters, functionality, and other properties (e.g., attack vulnerability) \cite{jagielski2020high}.

\noindent\textbf{Problem Settings.} MEAs represent a sophisticated class of threats aimed at replicating or approximating the functionality of a target machine learning model, typically deployed within a MLaaS platform. In this scenario, an adversary seeks to construct a substitute model $\mathcal{M}'$ that closely mimics the behavior of the target model $\mathcal{M}$.

\noindent\textbf{Attacker's Goals.} The objectives of MEAs span multiple levels of model compromise. At the most basic level, attackers may seek functionality replication, aiming to create a substitute model that produces similar outputs to the target model. More sophisticated goals include exact parameter recovery, where attackers attempt to extract the precise weights and biases of the target model. Architecture inference represents another critical objective, where attackers work to determine the target model's structure and hyperparameters. Beyond model characteristics, some attacks aim at training data reconstruction, attempting to recover samples from the original training dataset, which can lead to serious privacy breaches.

\noindent\textbf{Attacker's Background Knowledge.} The attack landscape spans varying degrees of prior knowledge about the target system. In black-box settings, attackers operate with minimal information about the model architecture, while gray-box scenarios provide partial knowledge of the model's structure or parameters. Knowledge of training data distribution can range from complete ignorance to access to similar datasets from the same domain. Understanding of the training process may include information about optimization algorithms and hyperparameters. Additionally, attackers might possess varying levels of information about the deployment environment, including details about the hosting platform and infrastructure.

\noindent\textbf{Attacker's Capabilities.} The practical constraints and resources available to attackers influence attack strategies. Query access represents the primary capability, where attackers can interact with $\mathcal{M}$ through API calls to obtain outputs $\mathcal{M}(\mathbf{x})$ for chosen inputs, typically subject to a query budget $B$ due to rate limiting or cost constraints. Computational resources $R$ and time constraints $T$ further bound the attack's scope. Some attackers may exploit side-channel access, observing non-functional characteristics $\mathcal{O}(\mathcal{M}, \mathbf{x})$ such as timing or power consumption patterns. The availability of substitute data for training the extracted model represents another crucial capability, whether the data follows similar or different distributions compared to the target's training data.

\subsection{Basics of Model Extraction Defenses}
\begin{figure}[htbp]
    \centering
    \includegraphics[width=\textwidth]{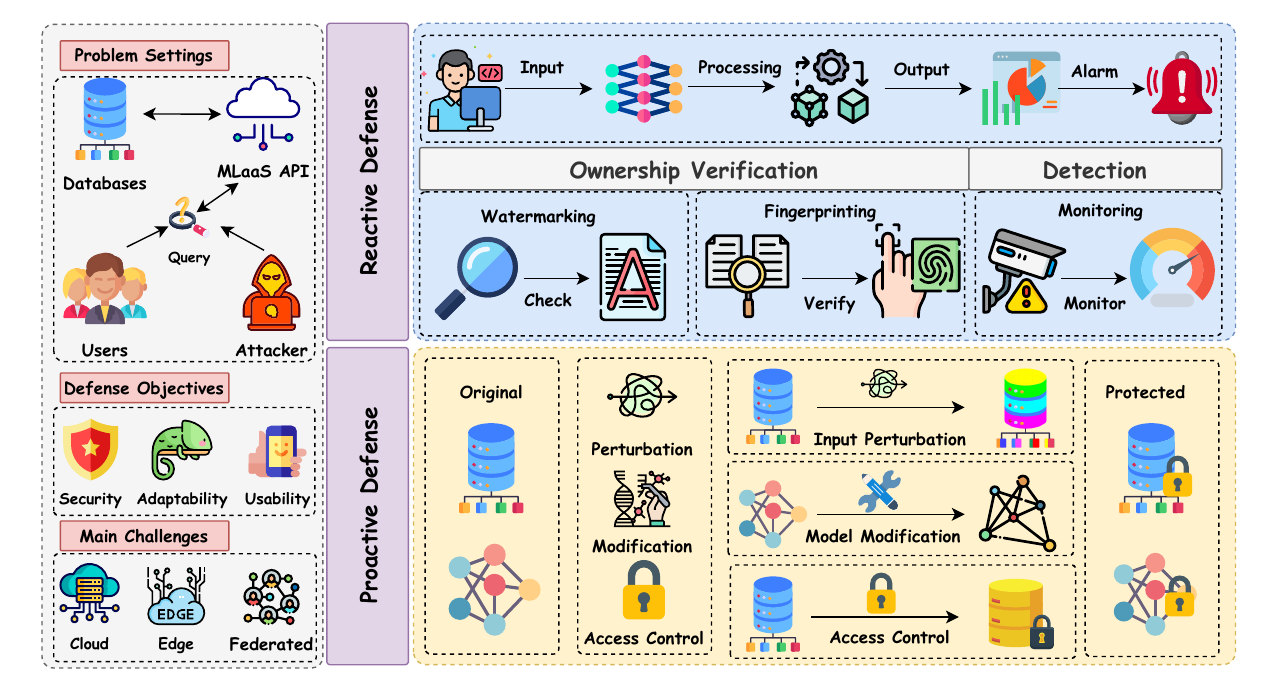}
    \caption{Model Extraction Defenses Overview. }
    \label{fig:med_overview}
\end{figure}

\noindent\textbf{Problem Setting.} MEDs represent protective measures designed to safeguard machine learning models deployed in Machine-Learning-as-a-Service (MLaaS) platforms against unauthorized replication attempts. These defenses aim to protect the intellectual property, privacy, and security of the target model while maintaining its utility for legitimate users.

\noindent\textbf{Defender's Goals.} The primary objectives of defenders span multiple aspects of model protection. The fundamental goal is to preserve model confidentiality by preventing unauthorized replication while maintaining model utility for legitimate users. This includes protecting model architecture, parameters, and training data from extraction attempts. Defenders also aim to maintain model performance under defensive measures, ensuring that protective mechanisms do not significantly degrade service quality. Additionally, they seek to establish verifiable ownership of their models and maintain robustness against evolving attack techniques.

\noindent\textbf{Defender's Background Knowledge.} Defenders operate with varying levels of information about potential threats and their own systems. This includes knowledge about the deployment environment, potential attack vectors, and user behavior patterns. Defenders may possess information about normal query patterns and usage statistics, enabling them to distinguish between legitimate use and potential attacks. Understanding of the model's sensitivity to different types of queries and knowledge about potential adversarial capabilities also influences defense strategy selection. This knowledge helps in calibrating defense mechanisms to balance security with usability.

\noindent\textbf{Defender's Capabilities.} The practical capabilities available to defenders shape their protection strategies. These include the ability to monitor and analyze query patterns, modify model responses, implement access controls, and deploy detection mechanisms. Defenders can typically control query responses, either through perturbation, truncation, or other transformation techniques. They may have the capability to implement rate limiting, user authentication, and logging mechanisms. Resource constraints such as computational overhead limits, latency requirements, and infrastructure capabilities bound the scope of deployable defenses. The ability to update and adapt defense mechanisms in response to new threats represents another crucial capability.

\section{Taxonomy of Model Extraction Attacks and Defenses} \label{taxonomy}

\begin{figure*}[ht]
\centering
\hspace*{-1.2cm}
\resizebox{1\textwidth}{!}{
\begin{forest}
  for tree={
    forked edges,
    edge={-, draw=black, line width=1.5pt},
    edge path={
      \noexpand\path[\forestoption{edge}, rounded corners]
      (!u.parent anchor) -- +(-1pt,0pt) -| ([xshift=0pt].child anchor)\forestoption{edge label};
    },
    grow=east,
    reversed,
    anchor=base west,
    parent anchor=east,
    child anchor=west,
    base=middle,
    font=\small\bfseries,
    rectangle,
    draw=black,
    line width=1pt,
    rounded corners,    
    align=left,
    minimum width=2em,
    s sep=5pt,
    l sep=0.5cm,
    inner sep=3pt,
  },
  where level=1{text width=4.5em}{},
  where level=2{text width=5em}{},
  where level=3{}{},
  where level=4{edge path={\noexpand\path[\forestoption{edge}] (!u.parent anchor) -- (.child anchor)\forestoption{edge label};}}{},
  where level=5{}{},
  [\textbf{Model Extraction}, fill=gray!20, align=center, anchor=north, edge=extraction, 
    [Attack, fill=red!20, edge=extraction, text width=3.5em
      [Query-based Attacks, fill=red!10, edge=extraction, text width=9.5em,
        [Substitute Model Training, fill=red!5, text width=11em, edge=extraction
          [\cite{rakin2022deepsteal, wang2022black, wu2022model, yuan2022es, sanyal2022towards, khaled2022careful, zhang2022text}, fill=white, text width=16.7em, edge=extraction]
        ]
        [Equation-solving Attacks, fill=red!5, text width=11em, edge=extraction
          [\cite{lee2022precise, dubey2022high, yoshida2019model}, fill=white, text width=16.7em, edge=extraction]
        ]
        [Recovery Attacks, fill=red!5, text width=11em, edge=extraction
          [\cite{truong2021data, li2021model, jia2021entangled, olatunji2023private}, fill=white, text width=16.7em, edge=extraction]
        ]
        [Meta-model Training, fill=red!5, text width=11em, edge=extraction
          [\cite{yoshida2024model, liu2024model, ren2024demistify, liu2024construct}, fill=white, text width=16.7em, edge=extraction]
        ]
        [Explanation-guided Attacks, fill=red!5, text width=11em, edge=extraction
          [\cite{yan2023explanation, oksuz2023autolycus, defazio2019adversarial}, fill=white, text width=16.7em, edge=extraction]
        ]
      ]
      [Data-driven Attacks, fill=red!10, edge=extraction, text width=9.5em,
        [Problem Domain, fill=red!5, text width=11em, edge=extraction
          [\cite{zhang2023ethicist, li2023model, yang2023efficient, oksuz2023autolycus, zhang2019data}, fill=white, text width=16.7em, edge=extraction]
        ]
        [Non-Problem Domain, fill=red!5, text width=11em, edge=extraction
          [\cite{zhu2023model, zhang2023plot, zhang2023survey, zhang2023apmsa, zhang2023extracting}, fill=white, text width=16.7em, edge=extraction]
        ]
        [Data-free Attack, fill=red!5, text width=11em, edge=extraction
          [\cite{sanyal2022towards, gurve2024misguide, miura2021megex, li2022data, truong2021data}, fill=white, text width=16.7em, edge=extraction]
        ]
      ]
      [Side-channel Attacks, fill=red!10, edge=extraction, text width=9.5em,
        [Software Side-channels, fill=red!5, text width=11em, edge=extraction
          [\cite{zhang2023apmsa, lee2022precise, zhang2021seat}, fill=white, text width=16.7em, edge=extraction]
        ]
        [Hardware Side-channels, fill=red!5, text width=11em, edge=extraction
          [\cite{breier2022sniff, dubey2022high, yoshida2019model}, fill=white, text width=16.7em, edge=extraction]
        ]
      ]
      [Gradient-based Attacks, fill=red!10, edge=extraction, text width=9.5em,
        [Direct Gradient Theft, fill=red!5, text width=11em, edge=extraction
          [\cite{li2023theoretical, tramer2017ensemble, papernot2018sok}, fill=white, text width=16.7em, edge=extraction]
        ]
        [Gradient Estimation, fill=red!5, text width=11em, edge=extraction
          [\cite{zhang2021graphmi, defazio2019adversarial, dai2018adversarial}, fill=white, text width=16.7em, edge=extraction]
        ]
      ]
      [Other Data Modalities, fill=red!10, edge=extraction, text width=9.5em,
        [Attacks on Text Models, fill=red!5, text width=11em, edge=extraction
          [\cite{bai, zhang2023ethicist, krishna2020thieves, he2021model, sha2024prompt, guo2024cold, zhang2022text, nazari2024llm, huang2022are}, fill=white, text width=16.7em, edge=extraction]
        ]
        [Attacks on Vision Models, fill=red!5, text width=11em, edge=extraction
          [\cite{kariyappa, truong2021data, orekondy2019knockoff, szyller2023good, shen2024prompt, liang2022imitated, spingarn2024stealing, zhao2024fully}, fill=white, text width=16.7em, edge=extraction]
        ]
        [Attacks on Graph Models, fill=red!5, text width=11em, edge=extraction
          [\cite{wu2024link, podhajski2024efficient, guan2024large, guan2024realistic, zhu2023model, zhang2021graphmi, olatunji2023private, zhang2023extracting}, fill=white, text width=16.7em, edge=extraction]
        ]
      ]
    ]
    [Defense, fill=green!20, edge=inversion, text width=3.5em
      [Attack Detection, fill=green!10, edge=inversion, text width=9.5em, 
        [Monitoring-based Methods, fill=green!5, text width=11em, edge=inversion
          [\cite{juuti2019prada, zhang2021seat, liu2024model, pal2021stateful, liu2022seinspect, chen2024queen, yan2021monitoring}, fill=white, text width=16.7em, edge=inversion]
        ]
      ]
      [Ownership Verification, fill=green!10, edge=inversion, text width=9.5em, 
        [Watermarking Techniques, fill=green!5, text width=11em, edge=inversion
          [\cite{tang2023exposing, szyller2021dawn, chakraborty2022dynamarks, zhao2021watermarking, he2022cater, tan2023deep, bachina2024genie, pang2024adaptive, li2024not}, fill=white, text width=16.7em, edge=inversion]
        ]
        [Fingerprinting Methods, fill=green!5, text width=11em, edge=inversion
          [\cite{cao2021ipguard, lukas2019deep, peng2022fingerprinting, guan2022you, xu2024instructional}, fill=white, text width=16.7em, edge=inversion]
        ]
      ]
      [Attack Prevention, fill=green!10, edge=inversion, text width=9.5em,
        [Data Perturbation, fill=green!5, text width=11em, edge=inversion
          [\cite{grana, kariyappa2020defending, zhou2024inversion, liang2024defending, zhang2023apmsa, lee2019defending, wu2024efficient}, fill=white, text width=16.7em, edge=inversion]
        ]
        [Model Modification, fill=green!5, text width=11em, edge=inversion
          [\cite{Szentannai2020preventing, li2021neurobfuscator, lin2020bident, xu2018deepobfuscation, goldstein2021preventing, sun2024streamlining, olney2022protecting, maungmaung2021protection}, fill=white, text width=16.7em, edge=inversion]
        ]
        [Access Control, fill=green!5, text width=11em, edge=inversion
          [\cite{dziedzic, kesarwani2018model}, fill=white, text width=16.7em, edge=inversion]
        ]
        [Adversarial Training, fill=green!5, text width=11em, edge=inversion
          [\cite{yilmaz, zhang2021seat,chen2021ast}, fill=white, text width=16.7em, edge=inversion]
        ]
      ]
      [Integrated Defense, fill=green!10, edge=inversion, text width=9.5em,
        [Holistic Methods, fill=green!5, text width=11em, edge=inversion
          [\cite{zhang2024defense,mori2021bodame,xian2022framework,jiang2023comprehensive,kariyappa2021protecting,li2024translinkguard,chabanne2020protection}, fill=white, text width=16.7em, edge=inversion]
        ]
        [Specialized Methods, fill=green!5, text width=11em, edge=inversion
          [\cite{wu2024efficient,liang2024defending,xie2024same,lee2022model,lin2020bident,guo2023isolation,maungmaung2021protection,mazeika2022steer}, fill=white, text width=16.7em, edge=inversion]
        ]
      ]
      [Other Data Modalities, fill=green!10, edge=inversion, text width=9.5em,
        [Defenses on Text Models, fill=green!5, text width=11em, edge=inversion
          [\cite{he2022cater,patil2023can,li2024translinkguard}, fill=white, text width=16.7em, edge=inversion]
        ]
        [Defenses on Vision Models, fill=green!5, text width=11em, edge=inversion
          [\cite{abuadbba2021deepisign, tang2023exposing,kariyappa2020defending,maungmaung2021protection,wu2024efficient}, fill=white, text width=16.7em, edge=inversion]
        ]
        [Defenses on Graph Models, fill=green!5, text width=11em, edge=inversion
          [\cite{zhao2021watermarking,waheed2023grove,you2024gnnguard,bachina2024genie,dai2024pregip}, fill=white, text width=16.7em, edge=inversion]
        ]
      ]
    ]
    [Different Computing Environments, fill=blue!20, edge=others, text width=15em
      [Cloud Computing, fill=blue!10, edge=others, text width=11em,
        [\cite{kesarwani2018model, gong2020model,yang2024swifttheft}, fill=white, text width=16.8em, edge=others, edge path={\noexpand\path[\forestoption{edge}] (!u.parent anchor) -- (.child anchor)\forestoption{edge label};}]
      ]
      [Edge Computing, fill=blue!10, edge=others, text width=11em,
        [\cite{kumar2021resource,rakin2022deepsteal,ren2024demistify,nazari2024llm,wu2023knowledge,meyers2024trained,nasr2019comprehensive,breier2022sniff,liu2024model}, fill=white, text width=16.8em, edge=others, edge path={\noexpand\path[\forestoption{edge}] (!u.parent anchor) -- (.child anchor)\forestoption{edge label};}]
      ]
      [Federated Learning, fill=blue!10, edge=others, text width=11em,
        [\cite{nasr2019comprehensive,zhu2019deep,zhao2020idlg,wang2019beyond,ganju2018property,qi2023differentially,mohassel2017secureml,abadi2016deep,geyer2017differentially}, fill=white, text width=16.8em, edge=others, edge path={\noexpand\path[\forestoption{edge}] (!u.parent anchor) -- (.child anchor)\forestoption{edge label};}]
      ]
    ]
  ]
\end{forest}
}
\vspace{-0mm}
\caption{A Taxonomy of Model Extraction Attacks \& Defenses on Machine Learning}
\vspace{-3mm}
\label{fig:taxonomy-techniques}
\end{figure*}
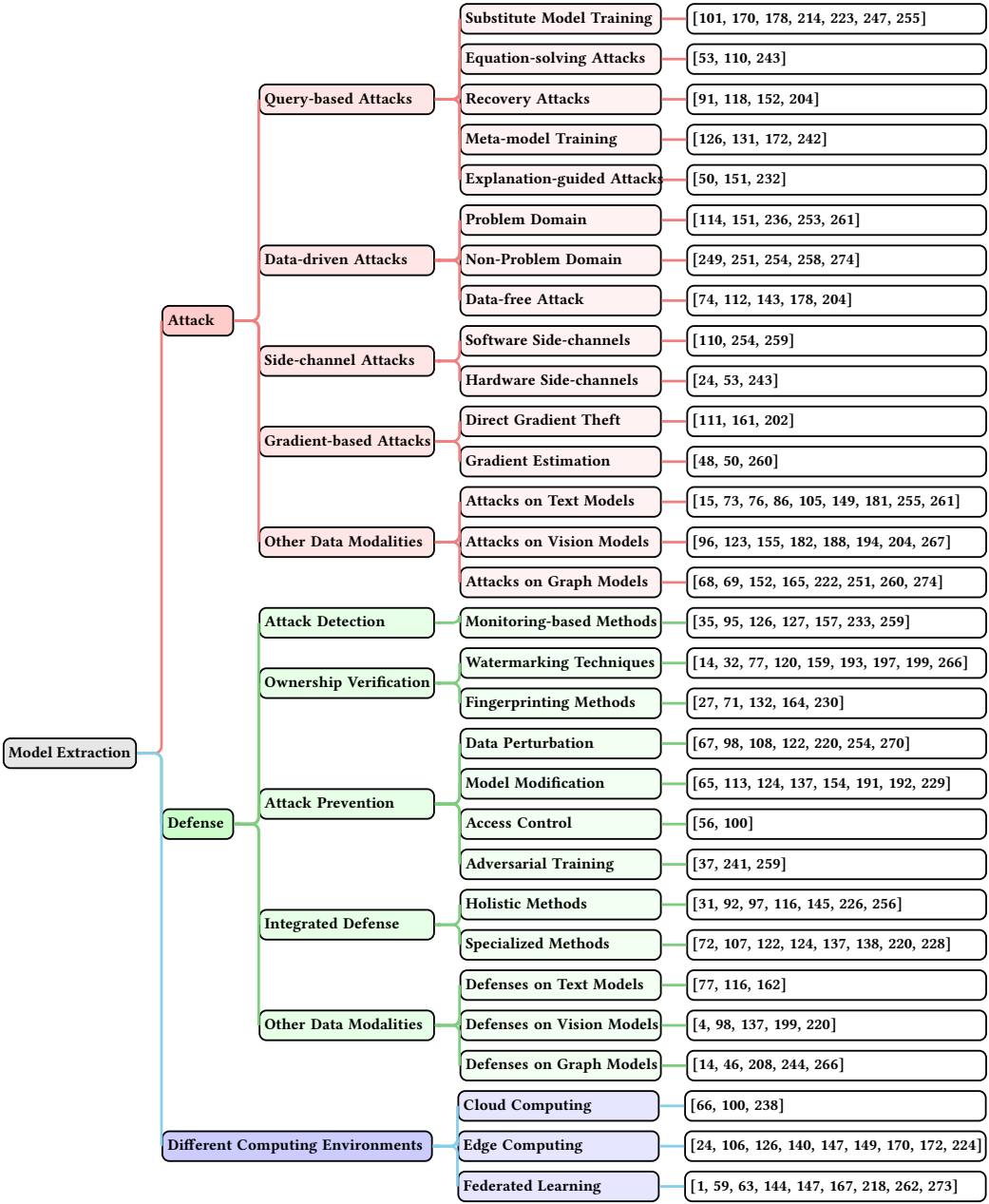

\noindent\textbf{Overview.} To give readers a general picture of MEAs and help readers find the most relevant papers easily, we
create a taxonomy of MEAs on ML models in Figure \ref{fig:taxonomy-techniques}. In this taxonomy, we systematically classify MEAs-related researches based on different attack types, defense strategies, and computing environments.

\noindent\textbf{Attack Classification.} In terms of attacks, we use the attack mechanism as the main categorization framework and subdivide it into five main categories: query-based attacks, data-driven attacks, side-channel attacks, gradient-based attacks, and attacks on other data modalities. For each attack type, we provide a detailed analysis in terms of several dimensions, including Technical Approaches, Target Models, and Attack Objectives. In particular, we also discuss the unique challenges of some of the attack types in specific Application Domains. We also pay special attention to attacks in special data modalities such as textual, visual and graph data modalities to highlight the unique challenges and solutions in these areas.

\noindent\textbf{Defense Strategies.} In terms of defense strategies, we divide them into four main categories: attack detection, ownership verification, attack prevention, and comprehensive and integrated defenses. We discuss in detail the technical implementation and challenges faced by each defense strategy. Also, we pay attention to special defense mechanisms under different data modalities (e.g., textual, visual, and graph structures) to reflect the unique needs of these domains.

\noindent\textbf{Attacks and Defenses in Different Computing Environments.} Finally, we separate out model extraction research in different computing environments, including cloud computing, edge computing, and federated learning environments. This section reflects the special techniques and challenges of defense in different computing paradigms, providing readers with a broader perspective.

\section{Model Extraction Attacks} \label{MEA}

Our taxonomy of MEAs is structured around the fundamental information channels through which model knowledge can be extracted, reflecting the natural progression from explicit to implicit extraction methods. At the most direct level, \textit{query-based attacks} exploit the essential input-output interface of deployed models, representing the primary and most straightforward channel for knowledge extraction. As access to this interface becomes restricted, attackers turn to \textit{data-driven approaches}, leveraging various data resources to replicate model behavior through more sophisticated learning techniques. When both query access and data resources are limited, \textit{side-channel attacks} emerge as a powerful alternative, exploiting unintended information leakage through physical and computational implementations. For models that expose or can be probed for gradient information, \textit{gradient-based attacks} provide a mathematically principled approach to extraction, targeting the optimization foundations of machine learning models. Finally, recognizing that \textit{different data modalities} present unique challenges and opportunities, our taxonomy addresses specialized attacks for text, vision, and graph models, capturing the distinct characteristics and vulnerabilities inherent in each domain.

\subsection{Query-based Attacks}

\subsubsection{Background of Query-based Attacks}
Query-based attacks represent a broad category of model extraction techniques that rely on systematically querying a target model to extract information about its functionality, parameters, or architecture \cite{tramer2016stealing, jagielski2020high}. These attacks are characterized by their fundamental dependence on the model's input-output behavior, where attackers iteratively present carefully crafted inputs to the target model and analyze its responses. The name "query-based" stems from the attack's core mechanism of repeatedly querying or interrogating the target model to gather information, analogous to how a scientist might probe a black box system to understand its internal workings. Within this category, several distinct approaches have emerged, including \textit{substitute model training} \cite{papernot2017practical, orekondy2019knockoff}, which creates substitute models mimicking target behavior; \textit{equation-solving attacks} \cite{reith2019efficiently, Wang2018StealingHI}, which treat extraction as a parameter recovery problem; \textit{recovery attacks} \cite{jagielski2020high, Milli2018ModelRF}, which directly target model components; \textit{meta-model training} \cite{oh2018towards}, which develops generalizable extraction strategies; and \textit{explanation-guided attacks} \cite{aivodji2020model, milli2019model}, which leverage model interpretability mechanisms. The effectiveness of these approaches varies depending on factors such as model complexity, prior knowledge, and practical constraints like query budgets and computational resources, though they all share the common foundation of systematic model interrogation.

\subsubsection{Substitute Model Training}
Substitute model training represents an approach to model extraction where an attacker constructs and trains a substitute model to replicate the behavior of a target model through strategic interaction \cite{tramer2016stealing, papernot2017practical, orekondy2019knockoff}. Unlike other extraction methods that focus on specific model components or characteristics, substitute model training aims to create a functionally equivalent copy of the entire target model by learning from its input-output behavior. This approach has been successfully demonstrated across various domains, from traditional machine learning models \cite{tramer2016stealing, reith2019efficiently} to complex deep learning architectures \cite{krishna2020thieves, takemura2020model, wu2022model}.
The training process involves two fundamental challenges: designing an appropriate model architecture and efficiently gathering training data through model queries. This leads to a natural organization of techniques into two main categories: \textit{architecture-driven approaches} and \textit{query optimization methods}.

\noindent
\textbf{Architecture Selection.}
The choice of substitute model architecture fundamentally determines the potential for successful extraction \cite{orekondy2019knockoff, shi2017how, juuti2019prada}. Modern approaches typically employ architectures of equal or greater capacity than the suspected target model structure, adapting to specific domains and tasks. For vision tasks, attackers often use sophisticated CNN architectures \cite{shi2017how, juuti2019prada, yu2020cloudleak}, while language model extraction requires transformer-based architectures \cite{krishna2020thieves, he2021model, takemura2020model}. Recent work has expanded this approach to more specialized models, including graph neural networks \cite{wu2022model, shen2021model}, GANs \cite{hu2021stealing, szyller2023good}, self-supervised learning models \cite{liu2022stolenencoder, sha2023cant, Yan2022TowardsEM}, and reinforcement learning systems \cite{chen2021stealing, defazio2019adversarial}.

\noindent
\textbf{Query Strategy Optimization.}
The efficiency of substitute model training heavily depends on the strategy used to query the target model \cite{chandrasekaran2018model, pal2020activethief, shi2018active}. The objective can be formalized as
\begin{equation}
Q = \mathop{\arg\max}_{Q \subset \mathcal{X}, |Q| \leq B} I(Q; \mathcal{M}),
\end{equation}
where $Q$ represents the query set, $B$ is the query budget, and $I(Q; \mathcal{M})$ measures information gain from queries. Modern approaches combine active learning \cite{pal2020activethief, li2018query} with data synthesis methods \cite{truong2021data, yuan2022es}. Some researchers leverage reinforcement learning \cite{orekondy2019knockoff, zhang2021thief} and evolutionary algorithms \cite{barbalau2020black} for query optimization, while others explore adversarial examples \cite{papernot2017practical, juuti2019prada, li2018query} to generate informative queries. The query optimization process can be further refined through
\begin{equation}
\mathcal{A}^* = \mathop{\arg\min}_{\mathcal{A}} \mathbb{E}_{Q \sim \mathcal{A}}[L(\mathcal{M}', \mathcal{M}; Q)],
\end{equation}
where $\mathcal{A}$ represents the query selection strategy and $L$ measures model behavioral differences.

Recent advances have significantly expanded the capabilities of substitute model training. Novel techniques incorporate model explanations \cite{aivodji2020model, Yan2022TowardsEM} and confidence score estimation \cite{wang2022black, wang2022enhance} to enhance extraction fidelity. Researchers have also developed methods for extracting GANs \cite{hu2021stealing, szyller2023good}, self-supervised learning models \cite{liu2022stolenencoder, sha2023cant}, and specialized architectures like graph neural networks \cite{wu2022model, He2020StealingLF}.
The effectiveness of substitute model training varies significantly across scenarios and target models. Architecture matching becomes critical for complex models, where CNNs require deep architectures \cite{orekondy2019knockoff, yu2020cloudleak} and transformer models demand attention mechanisms \cite{krishna2020thieves, he2021model}. Query strategies must adapt to specific constraints - active learning maximizes information gain with unlimited queries \cite{pal2020activethief, chandrasekaran2018model}, while synthesis methods like MAZE \cite{kariyappa} and other GAN-based approaches \cite{truong2021data, sanyal2022towards} enable attacks with limited data access. The success of extraction depends heavily on balancing these components against practical constraints like query budgets and architectural limitations, with different approaches showing distinct trade-offs between query efficiency, computational requirements, and extraction fidelity.

\subsubsection{Equation-solving Attacks}
Equation-solving attacks (ESAs) represent a class of model extraction techniques that formulate the extraction process as a system of equations, where the solution directly corresponds to the target model's parameters or hyperparameters. This approach is particularly effective when the extraction goal involves recovering exact values from the target model, such as learned parameters or training hyperparameters. The key characteristic of ESAs lies in their mathematical formulation, where adversaries construct and solve equation systems based on the model's input-output relationships or optimization conditions.

The fundamental principle of ESAs involves strategically querying the target model $\mathcal{M}$ to establish parameter constraints. Given a target model with parameters $\theta$, the attack can be formulated as
\begin{equation}
\text{Find } \theta \text{ such that } f_\theta(\mathbf{x}_i) = y_i \text{ for } i = 1, \ldots, n,
\end{equation}
where $f_\theta$ represents the target model's function, $\mathbf{x}_i$ are carefully selected input samples, and $y_i$ are the corresponding outputs obtained from querying $\mathcal{M}$. This mathematical framework provides the foundation for various parameter extraction scenarios.

\noindent
\textbf{Parameter Extraction for Linear Models.}
ESAs demonstrate particular effectiveness in extracting parameters from linear models, where the mathematical relationship between inputs and outputs can be explicitly formulated. For logistic regression, the equation system takes the form:
\begin{equation}
\sigma(\theta^\top \mathbf{x}_i) = y_i,
\end{equation}
where $\sigma$ denotes the logistic function. This formulation exploits the deterministic nature of linear models, allowing attackers to construct a well-defined system of equations. The process involves strategically selecting input points $\mathbf{x}_i$ that maximize the information gained about the model parameters $\theta$ with each query. This approach is remarkably efficient, typically requiring only 1-4 queries per parameter for exact recovery, significantly outperforming other extraction methods that might need hundreds or thousands of queries. The efficiency stems from the direct mathematical relationship between inputs and outputs in linear models, allowing each query to provide precise information about the target parameters. For support vector machines (SVMs) with linear kernels, \citet{reith2019efficiently} extended this approach by showing that parameters can be recovered through:
\begin{equation}
y_i(\mathbf{w}^\top \mathbf{x}_i + b) \geq 1 - \xi_i,
\end{equation}
where $\mathbf{w}$ and $b$ are the SVM parameters, and $\xi_i$ are slack variables. This formulation captures the geometric properties of SVMs, where the decision boundary is defined by support vectors. The attack exploits the fact that support vectors lie on the margin boundaries, allowing the attacker to reconstruct the hyperplane parameters through careful query selection.

\noindent
\textbf{Hyperparameter Recovery.}
ESAs extend beyond model parameters to extract training hyperparameters, an attack method that targets the fundamental training configuration of machine learning models. \citet{Wang2018StealingHI} demonstrated that regularization coefficients could be recovered by exploiting the optimality conditions of the model's objective function:
ESAs extend beyond model parameters to extract training hyperparameters, an attack method that targets the fundamental training configuration of machine learning models.\begin{equation}
\nabla_\theta L(\theta) + \lambda \nabla_\theta R(\theta) = 0,
\end{equation}
where $L$ represents the loss function, $R$ is the regularization term, and $\lambda$ is the regularization coefficient. This approach leverages a key insight from optimization theory: at the optimal solution, the gradient of the objective function must vanish. By analyzing the relationship between the loss gradient and regularization gradient, attackers can infer the relative importance assigned to the regularization term during training. This extraction technique is particularly significant because hyperparameters often encode critical design choices and domain expertise that organizations consider proprietary. The attack process involves constructing multiple gradient equations using different input points and solving the resulting system to determine the regularization coefficient. The success of this approach depends on having access to the model's architecture and parameters, which might require combining it with other extraction techniques. Despite this limitation, hyperparameter extraction represents a significant threat to model security, as these values often encode valuable information about the training process and data characteristics.

\noindent
\textbf{Applications and Limitations.}
Recent developments have shown ESAs' capability to circumvent privacy-preserving mechanisms. \citet{yan2022monitoring} successfully attacked models protected by differential privacy defenses through strategic query duplication and output analysis. However, ESAs face significant challenges with complex architectures, particularly deep neural networks, where equation systems become increasingly difficult to formulate and solve. Despite these limitations, ESAs remain a powerful threat to model confidentiality, especially for widely deployed simpler models like logistic regression and SVMs, where their precision and efficiency make them effective.

\subsubsection{Recovery Attacks}
Recovery attacks (RAs) represent a class of model extraction techniques that aim to precisely reconstruct the exact parameters and architecture of deep neural networks. RAs focus on recovering the actual weights and biases of the target model through careful analysis of its behavior. These attacks are particularly effective against networks using piecewise linear activation functions such as ReLU, where the relationship between inputs and outputs can be mathematically analyzed through geometric properties.

\noindent
\textbf{Parameter Recovery Fundamentals.}
The core principle of RAs involves analyzing the geometric properties of neural networks to reconstruct their parameters. \citet{Milli2018ModelRF} established the theoretical foundation by viewing network weights as hyperplanes in input space $\mathcal{X}$. This geometric interpretation enables weight recovery through careful identification of decision boundary points. For ReLU networks of arbitrary depth, \citet{Rolnick2019ReverseengineeringDR} formalized the extraction process as:
\begin{equation}
\text{For each layer } l: \text{ Find } \mathbf{W}_l, \mathbf{b}_l \text{ such that } h_l = \text{ReLU}(\mathbf{W}_l h_{l-1} + \mathbf{b}_l),
\end{equation}
where $h_l$ represents the layer output, and $\mathbf{W}_l$, $\mathbf{b}_l$ are the weight matrix and bias vector respectively. This formulation provides a systematic approach to recovering network parameters layer by layer, though the complexity increases significantly with network depth.

\noindent
\textbf{Advanced Extraction Methods.}
Building on these fundamentals, researchers have developed more sophisticated extraction techniques. \citet{Jagielski2019HighAA} demonstrated practical implementations achieving exact model replication, while \citet{Carlini2020CryptanalyticEO} introduced cryptanalytic perspectives to reduce query complexity. These advances have shown that RAs can achieve perfect model transferability under certain conditions. 

However, the practical application of RAs faces significant challenges, particularly in terms of computational requirements. For instance, extracting parameters from models with $2^{10}$ parameters typically requires $3^{12}$ queries per parameter, making the approach impractical for large-scale networks. Additionally, the effectiveness of RAs diminishes when targeting networks with non-piecewise linear activations or complex architectural features.
Despite these limitations, RAs remain particularly valuable in scenarios requiring exact model reproduction, such as security auditing or intellectual property verification. Their ability to recover exact parameters provides unique insights into model architecture and design choices, though this precision comes at the cost of increased computational complexity and query requirements. The trade-off between extraction accuracy and efficiency continues to drive research in this area, with ongoing efforts focused on developing more practical approaches while maintaining high fidelity in parameter recovery.

\subsubsection{Meta-model Training Attacks}
Meta-model training attacks represent an advanced model extraction approach that leverages meta-learning techniques to infer the architectural characteristics and hyperparameters of target models. Unlike direct parameter extraction methods, these attacks focus on learning the high-level design patterns and structural information of neural networks by training a meta-model to analyze the target model's behaviors and responses. A meta-model attack is defined when an adversary trains a separate model (meta-model) that takes as input the target model's responses to carefully crafted queries and outputs predictions about the target model's architecture, hyperparameters, or training configuration.

\noindent
\textbf{Meta-learning Framework.}
The foundational approach to meta-model attacks was established by \citet{oh2018towards}, who developed a systematic framework for inferring model architectures through query analysis. Their method trains a meta-model $\mathcal{M}'$ to predict architectural characteristics of a target model $\mathcal{M}$ using the following optimization objective:
\begin{equation}
\mathcal{M}' = \mathop{\arg\min}_{\mathcal{M}' \in \mathcal{H}} \mathbb{E}_{(\mathcal{M}, Q) \sim \mathcal{D}}[L(\mathcal{M}'(Q), \text{Arch}(\mathcal{M}))],
\end{equation}
where $\mathcal{H}$ represents the space of possible meta-models, $Q$ denotes a set of queries, $\mathcal{D}$ is the distribution over target models and queries, and $\text{Arch}(\mathcal{M})$ represents the target model's architectural details. This framework enables the prediction of various model characteristics, including layer configurations, activation functions, and optimization parameters. The approach demonstrates significant effectiveness, achieving high accuracy in hyperparameter prediction, though it requires substantial computational resources (typically 40 GPU days).

The success of meta-model training attacks depends heavily on several key factors. First, the diversity and representativeness of the training dataset used to train the meta-model significantly impact its ability to generalize to different target architectures. The meta-model must be exposed to a wide range of architectural patterns during training to effectively recognize similar patterns in unknown target models. Second, the quality and design of the query strategy play a crucial role - queries must be crafted to maximize the information gained about the target model's architecture while minimizing the number of required interactions. However, these attacks face limitations in scenarios where the target model's architecture significantly differs from those seen during meta-model training, or when the target model implements defensive measures that obscure its behavioral patterns.
While meta-model training attacks represent a sophisticated approach to model extraction, their practical deployment often requires significant computational resources and careful consideration of the trade-off between inference accuracy and efficiency. These attacks are valuable in scenarios where understanding a model's architectural decisions is more important than obtaining exact parameter values, such as in competitive analysis or security auditing.

\subsubsection{Explanation-guided Attacks}
Explanation-guided attacks represent a specialized class of model extraction techniques that exploit interpretability mechanisms in machine learning models. These attacks specifically target models equipped with explanation capabilities - systems designed to provide insights into their decision-making process through various forms of explanations. The attacks were first systematically studied by \citet{milli2019model}, who demonstrated how model explanations could be leveraged for efficient model reconstruction. An explanation-guided attack is defined when an adversary utilizes explanation outputs, beyond just model predictions, to enhance the extraction process, typically achieving higher fidelity and requiring fewer queries compared to traditional approaches.

The fundamental approach to explanation-guided attacks can be formalized through an optimization objective that considers both the model's predictions and its explanations:
\begin{equation}
\mathcal{M}' = \mathop{\arg\min}_{\mathcal{M}' \in \mathcal{H}} \mathbb{E}_{(x,y) \sim \mathcal{D}}[L(\mathcal{M}'(x), y) + \lambda D(E_{\mathcal{M}}(x), E_{\mathcal{M}'}(x))],
\end{equation}
where $\mathcal{M}'$ represents the extracted model, $\mathcal{H}$ is the hypothesis space, $L$ is a loss function measuring prediction accuracy, $E_{\mathcal{M}}(x)$ and $E_{\mathcal{M}'}(x)$ are the explanations from the target and extracted models respectively, and $D$ measures the distance between explanations. This formulation captures the dual objectives of matching both the model's predictions and its explanation patterns.

Recent research has demonstrated various sophisticated approaches to leveraging model explanations. \citet{miura2021megex} introduced MEGEX, which achieves efficient extraction in data-free settings by utilizing explanation information. \citet{oksuz2023autolycus} developed AUTOLYCUS, which combines feature importance scores and other explanatory information to guide synthetic sample generation, significantly improving extraction efficiency. In a different direction, \citet{Wang2022DualCFEM} proposed DualCF, which leverages counterfactual explanations to accurately reconstruct model decision boundaries, addressing previous limitations in boundary accuracy.

These advances have shown that explanation-guided attacks can significantly enhance extraction effectiveness across different scenarios. These attacks prove particularly powerful when explanations reveal information about the model's internal processing, such as feature attribution patterns or decision rationales, enabling attackers to gain deeper insights into the target model's decision-making process while requiring fewer queries. However, their effectiveness depends heavily on the quality and type of available explanations, as well as the computational resources required to process and utilize this additional information effectively.

\subsection{Data-driven Attacks}
Data-driven attacks encompass a broad category of model extraction techniques characterized by their use of data to query and replicate target models. These attacks can be classified based on the relationship between the data used for extraction and the target model's training domain. A data-driven attack is defined as an approach where an adversary systematically queries a target model using carefully selected or generated data points to reconstruct its functionality. The effectiveness and methodology of these attacks vary significantly depending on data availability and domain alignment, leading to three distinct categories: \textit{problem domain attacks, non-problem domain attacks, and data-free attacks}.

\subsubsection{Problem Domain Attacks}
Problem domain attacks represent the most straightforward approach to model extraction, where attackers have access to data from the same or similar distribution as the target model's training data. These attacks leverage the inherent similarity between the available data and the target model's training domain to achieve high-fidelity extraction. The mathematical formulation for these attacks can be expressed as:
\begin{equation}
\mathcal{M}' = \mathop{\arg\min}_{\mathcal{M}' \in \mathcal{H}} \mathbb{E}_{(x,y) \sim \mathcal{D}_{\text{PD}}}[L(\mathcal{M}'(x), \mathcal{M}(x)) + \lambda R(\mathcal{M}')],
\end{equation}
where $\mathcal{M}'$ is the extracted model, $\mathcal{H}$ is the hypothesis space, $\mathcal{D}_{\text{PD}}$ represents the problem domain data distribution, $L$ is a loss function comparing model outputs, and $R$ is a regularization term.

\noindent
\textbf{Transfer Learning and Active Sampling.}
The effectiveness of problem domain attacks stems from two key mechanisms. First, transfer learning approaches, exemplified by Knockoff Nets \citep{orekondy2019knockoff}, leverage pre-trained models and domain-specific features to accelerate extraction. This approach has shown particular success with complex architectures like deep neural networks and language models \citep{krishna2020thieves}. Second, active learning strategies, demonstrated by ActiveThief \citep{pal2020activethief}, optimize query selection to maximize information gain while minimizing queries. These methods identify and exploit the most informative regions of the input space, particularly effective when query access is limited or costly \citep{chandrasekaran2020exploring}.

\subsubsection{Non-problem Domain Attacks}
Non-problem domain (NPD) attacks operate in a more challenging scenario where attackers must use data from a different distribution than the target model's training domain. These attacks bridge the domain gap while extracting model functionality. The mathematical framework can be expressed as:
\begin{equation}
\mathcal{M}' = \mathop{\arg\min}_{\mathcal{M}' \in \mathcal{H}} \mathbb{E}_{x \sim \mathcal{D}_{\text{NPD}}}[L(\mathcal{M}'(x), \mathcal{M}(x)) + \lambda R(\mathcal{M}', \mathcal{D}_{\text{NPD}}, \mathcal{D}_{\text{PD}})],
\end{equation}
where $\mathcal{D}_{\text{NPD}}$ represents the non-problem domain data distribution, and $R$ includes domain shift.

\noindent
\textbf{Domain Adaptation Strategies.}
The key challenge in NPD attacks lies in bridging the distribution gap between available data and the target domain. Copycat CNN \citep{correia2018copycat} demonstrated that even random data can enable successful extraction through careful query selection. This approach was further refined by Knockoff Nets \citep{orekondy2019knockoff}, which introduced sophisticated domain adaptation techniques to align NPD data with the target domain. Active learning strategies, as shown in ActiveThief \citep{pal2020activethief}, complement these methods by intelligently selecting the most informative query points, particularly valuable when query access is restricted.

\subsubsection{Data-free Attacks}
Data-free attacks represent the most challenging scenario, where attackers must extract model functionality without access to any real data. These attacks rely on synthetic data generation and sophisticated optimization techniques. The formulation is:
\begin{equation}
\min_{\mathcal{M}', G} \max_{D} \mathbb{E}_{z \sim p(z)}[L(\mathcal{M}'(G(z)), \mathcal{M}(G(z))) + \lambda D(G(z))],
\end{equation}
where $G$ generates synthetic samples, $D$ encourages sample generation, and $z$ is a latent vector.

\noindent
\textbf{Synthetic Data Generation and Optimization.}
Data-free attacks primarily rely on two key mechanisms. First, adversarial generation methods, exemplified by MAZE \citep{kariyappa} and DFME \citep{truong2021data}, create synthetic samples that maximize information gain about the target model. Second, zeroth-order optimization techniques, developed in works like DFMS-HL \citep{sanyal2022towards}, enable effective model extraction even without gradient information. These approaches have been further enhanced by leveraging model explanations, as demonstrated by MEGEX \citep{miura2021megex}, which exploits additional information from explainable AI systems.

The three categories of data-driven attacks present distinct trade-offs in terms of effectiveness, efficiency, and practicality. Problem domain attacks typically achieve the highest fidelity but require access to similar data as the target model. NPD attacks offer greater flexibility but face challenges in bridging domain gaps, often requiring more queries or sophisticated adaptation techniques. Data-free attacks provide the most general approach but typically require more computational resources and may achieve lower fidelity.
The choice of attack strategy depends on several key factors: (1) Data availability: Problem domain attacks require similar data, while NPD and data-free attacks operate with less or no relevant data. (2) Query access: Active learning strategies in both problem and non-problem domain attacks can optimize limited query budgets, while data-free methods often require more queries. (3) Computational resources: Data-free attacks, particularly those using adversarial generation, demand significant computational power compared to data-driven approaches. (4) Target model characteristics: The effectiveness of specific techniques varies with model architecture and task domain, as demonstrated by specialized approaches for CNNs \citep{orekondy2019knockoff}, language models \citep{krishna2020thieves}, and GNNs \citep{wu2022model}.
Each category faces distinct limitations: Problem domain attacks may be impractical when similar data is unavailable; NPD attacks must overcome domain shift, potentially reducing extraction fidelity; and data-free attacks often require more queries and computational resources while potentially achieving lower accuracy. Understanding these trade-offs is crucial for both attackers in selecting appropriate strategies and defenders in developing comprehensive protection mechanisms.

\subsection{Side-channel Attacks}

\subsubsection{Background on Side-channel Attacks}
Side-channel attacks (SCAs) represent a class of model extraction techniques that exploit unintended information leakage from the physical implementation of machine learning systems rather than directly querying the model interface \citep{standaert2010introduction}. Side-channel refers to any observable physical or logical effect that provides information about the model's execution beyond its intended inputs and outputs, such as timing patterns, power consumption, electromagnetic emissions, or cache access patterns. Originally developed for cryptographic key recovery \citep{joy2011side}, side-channel attacks have evolved to target machine learning models, leveraging the fact that physical implementations inevitably leak information through measurable effects during computation. These attacks are particularly powerful because they can bypass traditional security measures that focus on protecting the model's interface, potentially revealing sensitive information about model architecture, parameters, and operations through careful measurement and analysis of physical phenomena. The effectiveness of side-channel attacks depends heavily on the attacker's ability to measure and interpret these unintended information channels, making them especially concerning in scenarios where adversaries have physical access to the hardware executing the model.
\subsubsection{Software Side-channels}
Software side-channel attacks represent a class of model extraction techniques that exploit information leakage through software-level observations of model execution. These attacks leverage the insight that computational patterns during model inference can reveal significant information about the model's architecture and parameters. The foundational principle can be formalized as:
\begin{equation}
\mathcal{M}' = \mathop{\arg\min}{\mathcal{M}' \in \mathcal{H}} \mathbb{E}_{x \sim \mathcal{X}}[D(S(\mathcal{M}(x)), S(\mathcal{M}'(x)))],
\end{equation}
where $\mathcal{M}'$ represents the extracted model, $\mathcal{H}$ is the hypothesis space, $S(·)$ captures observable side-channel information, and $D$ measures the distance between side-channel patterns.

\noindent
\textbf{Cache and Memory Analysis.}
The primary vector for software side-channel attacks exploits cache and memory access patterns during model execution \cite{hong2018security, Yan2020cache}. Cache-based techniques like Flush+Reload and Prime+Probe observe the shared cache architecture to infer model operations, while memory access pattern analysis \cite{liu2020ganred,Hu2020DeepSniffer} examines the broader sequence of memory operations. These approaches can reveal critical information about layer types, architecture depth, and operational patterns, making them particularly effective for extracting complex neural network.

\noindent
\textbf{Execution Timing Analysis.}
Timing analysis \cite{duddu2019stealing} exploits the temporal signatures of model execution to infer architectural details. This approach analyzes variations in execution time across different inputs to deduce model structure and complexity. GPU context-switching analysis \cite{hunt2020telekine} extends this concept to GPU-accelerated systems, examining kernel execution patterns to extract model information.

The effectiveness of software side-channel attacks varies with deployment context and target model characteristics. Cache-based approaches excel in shared computing environments but require local access to the target system. Timing analysis offers broader applicability but may suffer from noise in distributed settings. The choice of attack vector depends on factors like system access level, target model complexity, and deployment environment. While these attacks can effectively extract model information, their success is often constrained by system noise, virtualization barriers, and defensive countermeasures that obscure execution patterns.

\subsubsection{Hardware Side-channels}
Hardware side-channel attacks represent model extraction techniques that exploit physical characteristics of computing hardware during model execution. Unlike software side-channels that observe computational patterns, hardware side-channels measure physical phenomena such as power consumption, electromagnetic emissions, and hardware faults to infer model properties. This approach can be formalized as:
\begin{equation}
\mathcal{M}' = \mathop{\arg\min}{\mathcal{M}' \in \mathcal{H}} \mathbb{E}_{x \sim \mathcal{X}}[D(H(\mathcal{M}(x)), H(\mathcal{M}'(x)))],
\end{equation}
where $\mathcal{M}'$ represents the extracted model, $\mathcal{H}$ is the hypothesis space, $H(·)$ captures hardware side-channel measurements, and $D$ measures the distance between physical observations.

\noindent
\textbf{Power and Electromagnetic Analysis.}
The primary vectors for hardware side-channel attacks involve analyzing power consumption patterns and electromagnetic (EM) emanations during model execution. Power analysis techniques \cite{xiang2020open, dubey2022high} exploit variations in power consumption to reveal model architecture and parameters, particularly effective in edge computing scenarios. EM analysis \cite{batina2019csi, yoshida2019model, yu2020deepem} leverages unintended electromagnetic signals to extract detailed information about model structure, including layer configurations and activation functions. These approaches have proven particularly effective against hardware accelerators and FPGA implementations, where physical measurements can reveal significant information about model operations.

\noindent
\textbf{Invasive Hardware Attacks.}
More aggressive extraction techniques involve direct interference with hardware operation. Memory access pattern exploitation \cite{hua2018reverse, rakin2022deepsteal, wang2022demystifying} targets hardware accelerator memory systems to reveal architectural details and parameters. Fault injection attacks \cite{breier2021sniff} deliberately introduce hardware faults to extract model information, particularly effective against transfer-learned models. These approaches require physical access but can extract detailed model information even from secured systems.

The effectiveness of hardware side-channel attacks varies significantly with deployment context and hardware platform. Power and EM analysis provide comprehensive extraction capabilities but typically require physical proximity to the target device. Invasive techniques offer more detailed information but demand direct hardware access and risk device damage. These attacks are particularly concerning for edge devices and hardware accelerators, where physical security is often limited. Success depends on factors like hardware accessibility, measurement precision, and environmental noise, highlighting the need for comprehensive hardware security measures.

\subsection{Gradient-based Attacks}
\subsubsection{Background}
Gradient-based attacks represent a class of model extraction techniques that exploit gradient information - either directly accessible or estimated - to reconstruct model parameters, architecture, or training data. Unlike query-based or side-channel approaches that observe model outputs or physical characteristics, gradient-based attacks leverage the rich information contained in model gradients, which capture how model parameters respond to changes in inputs. These attacks can be formalized as optimization problems:
\begin{equation}
\min_{\theta'} D(\nabla_{\theta} L(\theta, x), \nabla_{\theta'} L(\theta', x))
\end{equation}
where $\theta$ and $\theta'$ represent target and extracted model parameters respectively, $\nabla_{\theta} L$ denotes parameter gradients, and $D$ measures gradient similarity. Gradient-based attacks typically fall into two main categories: direct gradient theft, which exploits explicitly available gradients (e.g., in federated learning), and gradient estimation, which approximates gradients through careful probing of black-box models. The effectiveness of these attacks stems from the fact that gradients contain comprehensive information about model architecture, parameters, and even training data, making them particularly powerful vectors for model extraction.
\subsubsection{Direct Gradient Theft}
Direct gradient theft attacks exploit explicitly available gradient information to extract model parameters or reconstruct training data. These attacks are particularly relevant in collaborative learning scenarios where gradients are shared among participants. The extraction process can be formalized as:
\begin{equation}
{\mathcal{M}', \mathcal{S}'} = \mathop{\arg\min}_{\mathcal{M}' \in \mathcal{H}, \mathcal{S}' \subset \mathcal{X}} D(\nabla L(\mathcal{M}, \mathcal{S}), \nabla L(\mathcal{M}', \mathcal{S}')),
\end{equation}
where $\mathcal{M}'$ represents the extracted model, $\mathcal{S}'$ the reconstructed dataset, and $D$ is gradient similarity.

\noindent
\textbf{Model Parameter Extraction.}
The primary approach in direct gradient theft focuses on reconstructing model parameters from gradient information. \citet{chen2023ddae} demonstrated that gradients shared during training can reveal substantial information about model architecture and parameters, developing D-DAE as a defense-penetrating extraction technique. \citet{karmakar2023marich} extended this concept with Marich, establishing theoretical foundations for query-efficient extraction attacks. These methods are particularly effective in distributed learning settings where gradient information is necessarily shared among participants.

\noindent
\textbf{Training Data Recovery.}
Another critical vector exploits gradient information to reconstruct training data, particularly concerning in privacy-sensitive applications. \citet{wainakh2022user} showed how gradient analysis could lead to user-level label leakage in federated learning, while \citet{boenisch2023when} demonstrated comprehensive training data reconstruction from aggregated updates. In language models, \citet{zhang2022text} revealed how gradient information could expose private text data from transformer architectures. \citet{li2023model} and \citet{li2023theoretical} further explored these vulnerabilities in distributed learning settings, providing theoretical frameworks for understanding and addressing gradient leakage risks.

The effectiveness of direct gradient theft varies with the learning context and available gradient information. Parameter extraction proves most powerful in collaborative learning environments where gradients are explicitly shared, while data recovery attacks become particularly concerning in privacy-sensitive applications. These attacks highlight the fundamental tension between model collaboration and privacy preservation in distributed learning systems.

\subsubsection{Gradient Estimation.}
Gradient estimation attacks enable model extraction in black-box scenarios where direct gradient access is unavailable. These attacks approximate gradient information through manipulation of model inputs and outputs. The estimation process can be formalized as:
\begin{equation}
\mathcal{M}' = \mathop{\arg\min}_{\mathcal{M}' \in \mathcal{H}} \mathbb{E}_{x \sim \mathcal{X}}[L(\mathcal{M}'(x), \mathcal{M}(x)) + \lambda R(\hat{\nabla} L(\mathcal{M}, x), \hat{\nabla} L(\mathcal{M}', x))],
\end{equation}
where $\hat{\nabla} L$ represents estimated gradients and $R$ measures the consistency of gradient estimates between target and extracted models.

\noindent
\textbf{Query-based Gradient Approximation.}
The primary approach uses systematic querying to approximate gradients. \citet{kariyappa} introduced MAZE, which employs zeroth-order optimization to estimate gradients through random perturbations:
\begin{equation}
\hat{\nabla} L(\mathcal{M}, x) \approx \frac{1}{N} \sum_{i=1}^N \frac{L(\mathcal{M}(x + \epsilon u_i)) - L(\mathcal{M}(x))}{\epsilon} u_i,
\end{equation}
where $u_i$ are random directions and $\epsilon$ controls perturbation size. \citet{chen2023ddae} extended this concept with finite difference methods in their D-DAE framework, demonstrating successful extraction even in the presence of defensive measures.

\noindent
\textbf{Adaptive Estimation Strategies.}
More sophisticated approaches dynamically adjust estimation strategies based on model responses. \citet{yang2024swifttheft} developed SwiftTheft, optimizing query patterns for efficient extraction from cloud-based models, while \citet{gao2024augsteal} proposed AugSteal, combining gradient estimation with data augmentation techniques. These methods demonstrate how adaptive sampling can significantly improve extraction efficiency under practical constraints like query budgets and time limitations.

The success of gradient estimation attacks depends heavily on query access and computational resources. Query-based approximation offers robust extraction capabilities but requires substantial queries, while adaptive strategies improve efficiency at the cost of increased complexity. These approaches reveal that even without direct gradient access, determined adversaries can extract significant model information through careful estimation techniques.

\subsection{Attacks Tailored for Specific Data Modalities}

\subsubsection{Attacks on Text Models}

Attacks on text models have emerged as a critical area of research in the field of model extraction and privacy preservation. These attacks target a wide range of natural language processing (NLP) models, from traditional architectures like BERT to the more recent large language models (LLMs). The unique characteristics of text data and the complex nature of language understanding make these models particularly vulnerable to various extraction techniques. We can broadly categorize these attacks into four main types: \textit{training data extraction, model functionality replication,  prompt engineering attacks, and fingerprinting attacks.}

\noindent\textbf{Training Data Extraction.} Training data extraction attacks aim to recover sensitive or proprietary information from the model's training corpus. These attacks exploit the tendency of language models to memorize and potentially reproduce parts of their training data. \citet{bai} introduced a novel approach using special characters to induce LLMs into revealing sensitive information. This method leverages the models' pattern completion capabilities, even when those patterns are crafted with non-standard characters. Similarly, \citet{zhang2023ethicist} proposed the "Ethicist" technique, combining loss smoothed soft prompting with calibrated confidence estimation to perform targeted extraction of specific information pieces. These attacks highlight a fundamental tension in language model design: the trade-off between model performance, which often improves with increased memorization, and data privacy. 
The potential for LLMs to leak personal information, as investigated by \citet{huang2022are}, underscores a fundamental challenge in training large-scale language models on diverse datasets. The inadvertent memorization and potential leakage of personal information from training data has significant implications for privacy and data protection.
As models grow larger and more sophisticated, they may inadvertently become more vulnerable to extraction attacks. This raises important questions about the ethical implications of training large language models on diverse datasets that may contain personal or copyrighted information.

\noindent\textbf{Model Functionality Replication.} Model functionality replication attacks focus on recreating the behavior of a target model, essentially stealing its intellectual property. \citet{krishna2020thieves} demonstrated the vulnerability of BERT-based APIs to such attacks, even with limited access. Their work showed that it's possible to create high-fidelity copies of these models, potentially compromising the competitive advantage of model owners. Building on this, \citet{he2021model} explored combined attacks that simultaneously extract model functionality and infer sensitive attributes. 
The unique structure of transformer-based models has also opened up new avenues for attacks. \citet{zhang2022text} demonstrated the possibility of reconstructing private text data from transformer models. This technique leverages the self-attention mechanisms and layer structures characteristic of transformers to infer information about the training data.
These attacks highlight the dual nature of the threat: not only can an attacker replicate the model's behavior, but they can also potentially extract sensitive information about the model's training data or architecture. This dual threat poses significant challenges for model developers, as protecting against one type of attack may not necessarily safeguard against the other.

\noindent\textbf{Prompt Engineering.}The advent of large language models has given rise to a new category of attacks centered around prompt engineering. \citet{sha2024prompt} introduced prompt stealing attacks that can reconstruct proprietary prompts used in downstream applications. This type of attack could undermine the competitive advantage of businesses relying on carefully crafted prompts for their LLM-based services. In a related vein, \citet{guo2024cold} proposed the COLD-Attack, a method for jailbreaking LLMs with stealthiness and controllability. These attacks exploit the very feature that makes LLMs so powerful: their ability to understand and generate human-like text based on prompts. The success of these attacks suggests that the security of LLM-based applications may depend not just on the model itself, but also on the careful design and protection of the prompts used to interact with it. This adds a new layer of complexity to the security landscape of text models, as developers must now consider the security of their prompts as well as their models.

\noindent\textbf{Fingerprinting attacks.} LLM-FIN attack introduced by \citet{nazari2024llm}, present yet another avenue for compromising model security, particularly for edge-deployed language models. These attacks raise concerns about the privacy and uniqueness of model deployments, especially in resource-constrained environments. The ability to fingerprint models could lead to targeted attacks or unauthorized tracking of model usage, posing risks for both model owners and users.

Looking ahead, the landscape of attacks on text models is likely to evolve rapidly as these models grow in complexity and capability. Future research may focus on developing more sophisticated attack methods that can exploit the nuanced understanding of language exhibited by advanced models. Potential areas of exploration include attacks that combine multiple techniques, such as leveraging prompt engineering to enhance training data extraction, or using model inversion techniques in conjunction with fingerprinting attacks to create more powerful extraction methods. Additionally, as text models are increasingly deployed in critical applications such as healthcare, finance, and legal domains, the potential impact of successful attacks grows more severe. This may drive research towards developing attack methods specifically tailored to these high-stakes domains, potentially exploiting domain-specific knowledge or vulnerabilities.

\subsubsection{Attacks on Vision Models}
Attacks on vision models have emerged as a critical area of research in the field of model extraction and privacy preservation, targeting a wide range of visual processing systems from convolutional neural networks (CNNs) to more complex architectures like image-to-image translation models and text-to-image generators. These attacks exploit the unique characteristics of visual data and the complex nature of visual processing to extract sensitive information, replicate model functionality, or steal intellectual property. We can broadly categorize these attacks into four main types: \textit{data-free attacks, limited-data attacks, task-specific attacks, and efficient extraction methods.}

\noindent\textbf{Data-free Attacks.} Data-free attacks represent a particularly challenging and innovative approach to model extraction in the vision domain. These attacks aim to reconstruct the functionality of target models without access to any real training data, relying instead on carefully crafted synthetic inputs or optimization techniques. The MAZE attack \cite{kariyappa} and the data-free model extraction method proposed by \cite{truong2021data} exemplify this approach, using zeroth-order gradient estimation and generative adversarial networks (GANs) respectively to synthesize queries that maximize the discrepancy between the target and substitute models. These techniques not only demonstrate the vulnerability of vision models to extraction attempts even in the absence of real data but also highlight the potential for attackers to circumvent data access restrictions or privacy protections. The success of these attacks raises significant concerns about the security of deployed vision models, particularly in scenarios where data privacy is paramount or where the original training data is proprietary.

\noindent\textbf{Limited-data Attacks. }Limited-data attacks, while less extreme than their data-free counterparts, still pose a significant threat to vision model security by demonstrating effective extraction with minimal data requirements. The Knockoff Nets approach \cite{orekondy2019knockoff} and the Copycat CNN method \cite{correia2018copycat} showcase how attackers can leverage small amounts of random, non-labeled data or publicly available datasets to steal the functionality of black-box vision models. These attacks are particularly concerning because they illustrate the vulnerability of models to extraction attempts using data that may be significantly different from the original training distribution. This capability challenges the assumption that model security can be maintained simply by restricting access to domain-specific data. Furthermore, the efficiency of these attacks in terms of query complexity and data requirements makes them potentially more practical and scalable in real-world scenarios.

\noindent\textbf{Task-specific Attacks.}Task-specific attacks target particular types of vision models or applications, exploiting the unique characteristics of these systems to achieve more specialized forms of model extraction. Attacks against image-to-image translation models \cite{szyller2023good}, object detectors \cite{liang2022imitated}, and text-to-image generation models \cite{shen2024prompt} demonstrate the breadth of vulnerabilities across different vision tasks. These attacks often leverage task-specific knowledge or constraints to enhance their effectiveness. For example, the attack on image-to-image translation models exploits the structured nature of the input-output pairs in these systems, while attacks on text-to-image models focus on extracting the crucial prompt-to-image mapping that underlies their functionality. The diversity of these task-specific attacks underscores the need for security measures that are tailored to the specific characteristics and vulnerabilities of different vision model architectures and applications.

\noindent\textbf{Efficient Extraction Attacks.} A particularly innovative direction in vision model attacks is the development of highly efficient extraction methods that maximize the information gained from a minimal number of queries. The SuperPixel Sample Gradient Model Stealing technique \cite{zhao2024fully} and the single-query attack on image-to-image translation models \cite{spingarn2024stealing} represent the cutting edge of this approach. These methods demonstrate that significant information about a model's functionality can be extracted with few interactions, challenging traditional assumptions about the query complexity required for effective model extraction. The success of these efficient attacks raises questions about the fundamental limits of model privacy and the feasibility of protecting vision models in scenarios where even a small number of queries might be sufficient for extraction.

The evolution of attacks on vision models reveals several key trends and challenges in the field of model security. First, the increasing sophistication of data-free and limited-data attacks suggests that restricting access to training data or using proprietary datasets may not be sufficient to protect vision models from extraction attempts. This trend necessitates the development of new protection mechanisms that focus on the inherent properties of the models themselves rather than relying solely on data access controls.
Second, the success of task-specific attacks across a wide range of vision applications highlights the need for a more nuanced understanding of security vulnerabilities in different types of vision models. Generic protection mechanisms may not be sufficient to address the unique challenges posed by specialized architectures or task-specific constraints.
Finally, the emergence of highly efficient extraction methods with minimal query requirements suggests that traditional approaches to query monitoring or rate limiting may become increasingly ineffective. This development calls for new paradigms in model protection that can maintain security even in the face of extremely limited model interactions.
\subsubsection{Attacks on Graph Models}

Attacks on graph models have emerged as a significant area of research in the field of model extraction and privacy preservation, targeting Graph Neural Networks (GNNs) and other graph-based machine learning models. These attacks exploit the unique structural properties and relational information inherent in graph data, presenting multifaceted challenges to the security and privacy of graph-based systems. We can broadly categorize these attacks into four main types: \textit{link stealing, structure inference, node feature extraction, and inductive learning attacks}.

\noindent\textbf{Link Stealing Attacks.} Link stealing attacks represent a fundamental threat to graph-based models, aiming to reconstruct the edges or connections between nodes in the underlying graph structure. These attacks exploit the model's learned representations to infer the presence or absence of links, potentially compromising sensitive relationship information in social networks, molecular structures, or other graph-based datasets. Recent works have demonstrated the feasibility of link stealing attacks against various GNN architectures, including those designed for inductive learning tasks \cite{wu2024link, guan2024large, podhajski2024efficient}. The success of these attacks highlights the vulnerability of graph models to privacy breaches, even when the original graph structure is not directly accessible. Moreover, the use of large language models as tools for link stealing, as proposed by \cite{guan2024large}, introduces a new dimension to these attacks, leveraging the powerful generalization capabilities of LLMs to enhance the effectiveness of link reconstruction. This trend towards incorporating advanced AI techniques in attack strategies signals a significant escalation in the sophistication of graph model attacks, potentially making them more challenging to detect and mitigate.

\noindent\textbf{Structure Inference Attacks.} Structure inference attacks extend beyond link stealing, attempting to reconstruct the overall topology of the graph or infer global structural properties. These attacks often leverage side information or multiple query responses to piece together a comprehensive view of the graph structure. Works such as \cite{guan2024realistic, zhu2023model} have demonstrated that it is possible to extract significant structural information from GNNs, even when access is limited to model predictions or embeddings. The success of these attacks raises substantial concerns about the confidentiality of graph structures in sensitive applications, such as social network analysis or molecular design, where the structure itself may be proprietary or contain sensitive information. The ability to infer global structural properties from limited local information highlights the inherent tension between the expressive power of graph-based models and their vulnerability to information leakage. This challenge is particularly acute in domains where the graph structure itself is the primary asset to be protected, necessitating the development of novel defense strategies that can preserve structural privacy without significantly compromising model utility.

\noindent\textbf{Node Feature Extraction.} Node feature extraction attacks focus on recovering the attribute information associated with individual nodes in the graph, exploiting the learned representations or model outputs to infer the original feature values. This type of attack potentially compromises personal or sensitive information encoded in node attributes. Research in this area, including works like \cite{zhang2021graphmi, olatunji2023private}, has shown that it is possible to extract meaningful feature information from GNNs, even when the models are designed with privacy-preserving mechanisms. The success of these attacks underscores the challenge of maintaining feature privacy in graph-based models, particularly in domains where node attributes may contain sensitive personal or proprietary information. The interdependence between node features and graph structure in many GNN architectures exacerbates this challenge, as protecting one aspect of the data may inadvertently expose vulnerabilities in another. This interconnectedness necessitates a holistic approach to graph model security that considers the interplay between structural and attribute information.

\noindent\textbf{Inductive Learning Attacks.} Inductive learning attacks represent a particularly challenging class of threats against graph models, targeting GNNs designed for generalization to unseen nodes or graphs. Works such as \cite{wu2024link, podhajski2024efficient} have demonstrated that it is possible to extract meaningful information or replicate model behavior even in these more challenging inductive settings. The success of these attacks highlights the need for robust defense mechanisms that can protect graph models across various learning paradigms and application scenarios. The ability of adversaries to generalize attacks to unseen data structures poses significant challenges for the deployment of graph-based models in dynamic or evolving environments, where the graph structure may change over time or where new nodes or subgraphs may be introduced. This vulnerability underscores the importance of developing adaptive defense strategies that can maintain security and privacy guarantees in the face of changing data distributions and model behaviors. \citet{wang2025cega} further study cost-constrained GNN model stealing and propose CEGA, which selects query nodes by representativeness, uncertainty, and diversity to achieve high fidelity under tight budgets.

The landscape of attacks on graph models is further complicated by the emergence of federated learning and other distributed computation paradigms in graph-based settings. Research such as \cite{chen2024empirical, zhang2023extracting} has shown that federated graph learning introduces new attack vulnerabilities and privacy concerns, as adversaries may attempt to extract sensitive information from the shared model updates or exploit the distributed nature of the computation to infer graph properties across multiple parties. These challenges are particularly acute in scenarios where graph data is distributed across multiple organizations or devices, each with potentially different privacy requirements and threat models. The tension between collaborative learning and data privacy in federated graph settings presents a unique set of challenges that require innovative solutions at the intersection of cryptography, distributed systems, and graph theory.

Recent work has also begun to explore the intersection of fairness and security in graph models. Studies such as \cite{zhang2024adversarial} have investigated adversarial attacks on the fairness of GNNs, highlighting how adversaries might manipulate model behavior to introduce or exacerbate biases in graph-based decision-making processes. This line of research underscores the complex interplay between security, privacy, and fairness in graph-based machine learning systems, illustrating how vulnerabilities in one aspect of model behavior can have far-reaching implications for the overall trustworthiness and ethical deployment of graph-based AI systems. The potential for adversarial attacks to manipulate model fairness raises important questions about the robustness of graph-based decision-making systems in high-stakes applications, necessitating the development of defense strategies that simultaneously address security, privacy, and fairness concerns.

The development of data-free MEAs against GNNs, as explored in \cite{zhuang2024unveiling}, represents a significant advancement in the field of graph model attacks. These attacks demonstrate that it is possible to extract meaningful information or replicate model behavior without access to the original training data, posing new challenges for the protection of graph-based models and intellectual property. The success of data-free extraction methods highlights the fundamental vulnerability of learned representations in graph models, suggesting that even carefully guarded training data may not be sufficient to protect against sophisticated extraction attempts. This development necessitates a reevaluation of traditional approaches to model protection and intellectual property in the context of graph-based machine learning, potentially requiring new paradigms for secure model deployment and knowledge sharing in graph-based AI systems.
The interplay between structural information, node features, and model behavior in graphs creates unique vulnerabilities that traditional machine learning security approaches may not adequately address. As graph models find applications in increasingly sensitive domains, from social network analysis to drug discovery, the potential impact of successful attacks grows more severe.

\section{Defenses against MEAs} \label{MED}

Our taxonomy of defense strategies against MEAs is organized based on the fundamental timing and approach of protection mechanisms. We identify four main categories that comprehensively cover defense strategies: attack detection, ownership verification, attack prevention, and comprehensive integrated frameworks. Detection methods focus on identifying extraction attempts during the querying phase. Ownership verification techniques, including watermarking and fingerprinting, provide post hoc attribution by verifying that a suspect model derives from a protected one. Prevention techniques, such as data perturbation, model modification, and query limitation, aim to proactively thwart attacks before they succeed by making the extraction process itself more challenging or less reliable. Comprehensive frameworks combine multiple defense mechanisms to provide holistic protection, addressing the multifaceted nature of extraction threats. This natural progression from detection and ownership verification to prevention and then integration reflects the evolution of defense strategies in response to increasingly attacks.

\subsection{Attack Detection}

\subsubsection{Monitoring-based Methods}
Monitoring-based methods for detecting MEAs can be broadly categorized into three main approaches: query pattern analysis, behavior anomaly detection, and information-theoretic monitoring. These methods aim to identify potential attacks by analyzing the interactions between users and the target model, leveraging various statistical and machine learning techniques to distinguish between benign and malicious behavior. The general formulation for monitoring-based methods can be expressed as
\begin{equation}
\mathcal{D}(Q, \mathcal{M}) = \begin{cases}
1 & \text{if } f(Q, \mathcal{M}) > \tau \\
0 & \text{otherwise}
\end{cases},
\end{equation}
where $\mathcal{D}$ is the detection function, $Q = {q_1, q_2, ..., q_n}$ is the set of queries made to the target model $\mathcal{M}$, $f$ is a monitoring function that analyzes the queries and model responses, and $\tau$ is a threshold for classifying behavior as malicious. The goal is to design $f$ such that it effectively distinguishes between benign queries and those indicative of model extraction attempts.

\noindent\textbf{Query Pattern Analysis:} This approach focuses on examining the patterns and distributions of queries made to the target model. One of the pioneering works in this category is PRADA (Protecting Against DNN Model Stealing Attacks) by \citet{juuti2019prada}, which analyzes the distribution of consecutive queries to detect abnormal patterns indicative of model extraction attempts. PRADA uses a statistical test based on the maximum mean discrepancy (MMD) to compare the distribution of queries from a user to a reference distribution of benign queries. Building upon this idea, \citet{zhang2021seat} proposed SEAT (Similarity Encoder by Adversarial Training), which employs a similarity encoder trained through adversarial learning to detect malicious queries. SEAT can capture more complex query patterns and is more robust against adaptive attackers who try to mimic benign query distributions. More recently, \citet{liu2024model} introduced a multi-dimensional feature-based detection method specifically designed for IoT services, which can improve detection accuracy in resource-constrained environments.

\noindent\textbf{Behavior Anomaly Detection:} This category of methods aims to identify unusual behavior patterns that may indicate ongoing MEAs. \citet{pal2021stateful} introduced a stateful detection approach that uses a modified variational autoencoder to track the distribution of queries over time, enabling the detection of subtle changes in user behavior that might be missed by stateless methods. Similarly, \citet{liu2022seinspect} proposed SeInspect, which employs a two-layer defense mechanism combining semantic feature analysis with user behavior monitoring. SeInspect's approach allows for efficient batch processing of user queries, reducing the latency overhead compared to traditional user-by-user inspection methods. These behavior-based approaches are particularly effective against attackers who carefully craft their queries to evade simple pattern-based detection methods.

\noindent\textbf{Information-Theoretic Monitoring:} A more recent trend in monitoring-based defenses involves leveraging information theory to quantify the amount of information leaked through model queries. \citet{chen2024queen} proposed QUEEN (Query Unlearning against Model Extraction), which uses a query unlearning mechanism to proactively reduce the information content of model responses. These information-theoretic approaches provide a more principled foundation for defense strategies, offering better theoretical guarantees on the protection they provide.

The effectiveness of monitoring-based methods varies significantly depending on the attack scenario and the characteristics of both the target model and its legitimate users. Query pattern analysis techniques like PRADA and SEAT excel at detecting attacks involving large numbers of structured queries, but may falter against highly adaptive attackers capable of mimicking benign query patterns. In contrast, behavior anomaly detection methods, such as those proposed by \citet{pal2021stateful} and \citet{liu2022seinspect}, show greater efficacy against sophisticated attackers but often require extended monitoring periods to establish accurate baseline behaviors. Information-theoretic approaches like ModelGuard and QUEEN offer robust theoretical guarantees at the cost of potentially higher computational overhead. A common challenge across all these methods is achieving an optimal balance between detection accuracy and false positive rates. Overly sensitive detectors risk flagging legitimate users as potential attackers, disrupting normal service operations. To mitigate this, some approaches, like that of \citet{yan2021monitoring}, incorporate differential privacy mechanisms to introduce controlled noise into the detection process, thereby reducing false positives while maintaining high detection rates for actual attacks. The effectiveness of these methods can be further enhanced by combining multiple techniques, as demonstrated by \citet{sadeghzadeh2023hoda} in their HODA framework, which integrates hardness-oriented features with traditional query pattern analysis for more robust detection. In conclusion, while monitoring-based methods offer a powerful and flexible approach to defending against MEAs without modifying the underlying model, the choice of specific technique must be carefully considered based on expected attack patterns, available computational resources, and the sensitivity of the protected model. The ongoing challenge lies in developing adaptive and efficient monitoring techniques that provide robust protection against diverse extraction attempts while minimizing impact on model performance.

\subsection{Ownership Verification}

\subsubsection{Watermarking Techniques}

Watermarking techniques have emerged as a prominent defense strategy against MEAs, offering a way to embed unique identifiers into machine learning models without significantly impacting their performance. These techniques can be broadly categorized into four main approaches: static watermarking, dynamic watermarking, domain-specific watermarking, and recent advancements in adaptive and robust watermarking. The general formulation for watermarking techniques can be expressed as
\begin{equation}
\mathcal{M}_w = g(\mathcal{M}, W, K),
\end{equation}
where $\mathcal{M}$ is the original model, $W$ is the watermark, $K$ is a secret key, $g$ is the watermarking function, and $\mathcal{M}_w$ is the watermarked model. The goal is to design $g$ such that
\begin{equation}
f_{\mathcal{M}w}(x) \approx f_{\mathcal{M}}(x) \quad \text{for all } x \in \mathcal{X},
\end{equation}
\begin{equation}
h(f_{\mathcal{M}_w}(x_w), y_w) = 1 \quad \text{for } (x_w, y_w) \in W,
\end{equation}
where $f_{\mathcal{M}}(x)$ and $f_{\mathcal{M}_w}(x)$ are the outputs of the original and watermarked models respectively, $\mathcal{X}$ is the input space, $h$ is a verification function, and $(x_w, y_w)$ are watermark trigger-response pairs. The watermark should be verifiable using $K$ while being difficult to remove or forge without knowledge of $K$.

\noindent\textbf{Static Watermarking.} Static watermarking methods involve embedding a fixed watermark into the model during training, which remains constant throughout the model's lifecycle. This approach is exemplified by the work of \citet{jia2021entangled}, who introduced the concept of entangled watermarks. Their method embeds watermarks by training the model on a mixture of original and watermarked data, creating a tight coupling between the model's performance and the presence of the watermark. This entanglement makes it difficult for attackers to remove the watermark without significantly degrading the model's performance. Similarly, \citet{tang2023exposing} proposed a robust and transferable watermark for thwarting MEAs, focusing on creating watermarks that are resistant to various extraction techniques.

\noindent\textbf{Dynamic Watermarking.} To address the limitations of static approaches, dynamic watermarking techniques have been developed. These methods allow for the watermark to be updated or changed over time, providing enhanced security against adaptive attacks. \citet{szyller2021dawn} proposed DAWN, a dynamic adversarial watermarking framework that continuously updates the watermark based on the model's interactions with potential attackers. This approach not only improves resilience against extraction attempts but also enables the detection of ongoing attacks. Similarly, \citet{chakraborty2022dynamarks} introduced DynaMarks, which uses dynamic watermarking to defend against deep learning model extraction. These dynamic approaches offer greater flexibility and adaptability compared to static methods, making them particularly suitable for scenarios where models are frequently updated or deployed in adversarial environments.

\noindent\textbf{Domain-Specific Watermarking.} Recognizing the unique challenges posed by different types of ML models, researchers have developed domain-specific watermarking techniques. In the realm of graph neural networks (GNNs), \citet{zhao2021watermarking} proposed a method for watermarking GNNs by embedding information into random graphs. This approach leverages the structural properties of graph data to create watermarks that are both effective and difficult to remove. For text generation models, \citet{he2022cater} introduced CATER, a conditional watermarking system specifically designed for text generation APIs. CATER embeds watermarks into the generated text in a way that is imperceptible to human readers but detectable by authorized parties. In the domain of deep neural networks, \citet{tan2023deep} presented a watermarking technique specifically designed to counter MEAs, tailoring the approach to the unique characteristics of deep architectures.

\noindent\textbf{Recent Advancements.} Recent research has focused on developing more sophisticated and robust watermarking techniques. \citet{bachina2024genie} proposed GENIE, a watermarking method for GNNs specifically designed for link prediction tasks, while \citet{dai2024pregip} introduced PreGIP, which watermarks the pretraining process of GNNs for enhanced intellectual property protection. \citet{pang2024adaptive} developed an adaptive and robust watermark specifically designed to counter MEAs, addressing the need for watermarks that can withstand sophisticated removal attempts. \citet{zhang2024defense} introduced a Bayesian active watermarking approach, combining watermarking with active learning principles to enhance detection capabilities. \citet{zhu2024reliable} focused on creating reliable model watermarks that defend against theft without compromising on evasion resistance, addressing the trade-off between protection and model performance. \citet{li2024not} proposed a multi-view data approach for watermarking deep neural networks, emphasizing the importance of learning feature-level watermarks rather than just modifying labels.

Watermarking techniques offer a promising defense against MEAs, with each approach having specific strengths and limitations. Static watermarking provides robust protection but may be vulnerable to long-term adaptive attacks, making it suitable for models with infrequent updates or in environments with lower security requirements. Dynamic watermarking addresses this limitation by offering adaptability and ongoing protection, making it ideal for frequently updated models or those deployed in highly adversarial settings. Domain-specific approaches are particularly effective for specialized models like GNNs or text generation systems, where the unique characteristics of the data or model architecture can be leveraged for more secure watermarking. Recent advancements in adaptive and robust watermarking techniques offer enhanced protection against sophisticated attacks but may require more computational resources or complex implementation. The choice of watermarking technique should be guided by the specific requirements of the model, the anticipated threat landscape, and the need to balance protection with performance. Despite their effectiveness, watermarking techniques face challenges such as potential watermark removal or overwriting by determined attackers, and the ongoing need to balance watermark robustness with model utility.

\subsubsection{Fingerprinting Techniques}

Fingerprinting techniques have emerged as a powerful approach to protect the intellectual property of machine learning models and verify their ownership. \citet{cao2021ipguard} pioneered this approach by introducing IPGuard, which creates fingerprints by exploiting the classification boundary of deep neural networks. Unlike watermarking, which often focuses on embedding information into model outputs, fingerprinting aims to create unique, verifiable patterns within the model itself. The general formulation for fingerprinting can be expressed as
\begin{equation}
\mathcal{M}_f = h(\mathcal{M}, F, K),
\end{equation}
where $\mathcal{M}$ is the original model, $F$ is the fingerprint, $K$ is a secret key, $h$ is the fingerprinting function, and $\mathcal{M}_f$ is the fingerprinted model. The goal is to design $h$ such that
\begin{equation}
f_{\mathcal{M}f}(x) \approx f_{\mathcal{M}}(x) \quad \text{for all } x \in \mathcal{X},
\end{equation}
\begin{equation}
v(f_{\mathcal{M}_f}(x_f), y_f, K) = 1 \quad \text{for } (x_f, y_f) \in F,
\end{equation}
where $f_{\mathcal{M}}(x)$ and $f_{\mathcal{M}_f}(x)$ are the outputs of the original and fingerprinted models respectively, $\mathcal{X}$ is the input space, $v$ is a verification function, and $(x_f, y_f)$ are fingerprint trigger-response pairs.

\noindent\textbf{Perturbation-based Fingerprinting.} One category of fingerprinting techniques leverages adversarial perturbations to create unique model behaviors. \citet{lukas2019deep} proposed a method using conferrable adversarial examples as fingerprints, which are designed to transfer with a target label from a source model to its substitutes. Extending this concept, \citet{peng2022fingerprinting} introduced a method based on Universal Adversarial Perturbations (UAPs), creating a unique profile of a model's decision boundary. \citet{tang2024modelguard} proposed ModelGuard, an output perturbation defense that uses information-theoretic principles to adaptively adjust perturbations to model outputs, minimizing information leakage while maintaining utility. These approaches have shown robustness against various MEAs and post-modification techniques, demonstrating good generalizability across different model architectures.

\noindent\textbf{Correlation-based Fingerprinting.} \citet{guan2022you} proposed a novel approach called Sample Correlation (SAC) for fingerprinting deep neural networks. This method leverages the pairwise relationship between sample outputs rather than point-wise indicators. SAC uses either wrongly classified normal samples (SAC-w) or CutMix augmented samples (SAC-m) to calculate correlation differences between models, offering robustness against various model stealing attacks, including adversarial training and transfer learning scenarios.

\noindent\textbf{Instruction-based Fingerprinting.} For large language models (LLMs), \citet{xu2024instructional} proposed Instructional fingerprinting, a method that uses lightweight instruction tuning to implant fingerprints. This technique creates a backdoor that causes the LLMs to generate specific outputs when presented with particular inputs. The approach proved effective across various LLMs architectures and demonstrated resilience to extensive fine-tuning.

From a technical perspective, each fingerprinting approach offers distinct advantages and trade-offs. Perturbation-based methods excel in models with well-defined decision boundaries, like image classifiers, but may struggle with discrete or high-dimensional output spaces typical in language models. Correlation-based techniques like SAC provide a balanced approach, offering robust detection across various architectures and attack scenarios, particularly where adversarial examples falter. Instruction-based methods are well-suited for complex models like LLMs, minimizing performance impact and resisting fine-tuning attacks, but require careful design to avoid task interference, especially in transfer learning contexts. The key challenge across all techniques lies in balancing fingerprint robustness with model utility, as overly aggressive fingerprinting can degrade performance, while subtle fingerprints risk removal or overwriting.

\subsection{Attack Prevention}

\subsubsection{Data Perturbation}
Data perturbation methods aim to defend against MEAs by introducing controlled noise or modifications to either the input data or output predictions of the target model. The general formulation for data perturbation defenses can be expressed as
\begin{equation}
\mathcal{D}(M(x)) = M(x) + \epsilon(x),
\end{equation}
where $M$ is the target model, $x$ is the input, $\mathcal{D}$ is the defense mechanism, and $\epsilon(x)$ is the perturbation added. The goal is to design $\epsilon(x)$ such that it significantly degrades the extraction attack while minimally impacting legitimate users.

\noindent\textbf{Input Perturbation.} Input perturbation techniques focus on modifying the query inputs before feeding them into the target model. The key insight behind these approaches is that MEAs often rely on querying the target model with out-of-distribution (OOD) or synthetic inputs. By selectively perturbing such inputs, defenders can disrupt the attacker's ability to gather accurate information about the model's decision boundaries.
\citet{grana} propose a method that adds small, carefully crafted perturbations to input queries detected as potential extraction attempts. These perturbations are designed to be imperceptible to humans but can significantly alter the model's predictions for malicious queries. Similarly, \citet{kariyappa2020defending} introduce an adaptive misinformation technique that identifies OOD queries and responds with incorrect predictions. This approach exploits the fact that legitimate users typically query the model with in-distribution samples, while attackers often use OOD data. By selectively providing misinformation for OOD queries, the defense can substantially degrade the accuracy of the attacker's clone model while maintaining high performance for benign users.
These input perturbation methods offer the advantage of being relatively lightweight and easy to implement as pre-processing steps. However, they may struggle with sophisticated attackers who can generate queries that closely mimic the distribution of legitimate inputs.

\noindent\textbf{Output Perturbation.} Output perturbation techniques focus on modifying the model's predictions or confidence scores before returning them to the querying party. These methods aim to obscure the true decision boundaries of the target model, making it more difficult for attackers to accurately replicate its behavior.
\citet{zhou2024inversion} propose an inversion-guided defense that detects potential model stealing attacks by analyzing the invertibility of the output predictions. When a suspicious query is detected, the defense mechanism perturbs the output to make it less informative for the attacker. \citet{liang2024defending} take a different approach by combining OOD feature learning with decision boundary confusion. Their method not only identifies OOD queries but also intentionally distorts the model's decision boundaries for such inputs, making it challenging for attackers to infer the true underlying structure of the target model.
Output perturbation techniques have the advantage of being able to adapt dynamically to different types of queries and attack patterns. However, they may introduce a trade-off between security and utility, as excessive perturbation can potentially impact the model's performance for legitimate users.

\noindent\textbf{Adversarial Perturbation.} Some defense methods leverage adversarial perturbation techniques, originally developed for robustness against adversarial examples, to counter MEAs. \citet{zhang2023apmsa} propose APMSA, which generates adversarial perturbations specifically designed to mislead potential model extraction attempts. This approach aims to proactively alter the model's behavior in ways that are difficult for attackers to reverse-engineer.
\citet{lee2019defending} introduce a defense strategy using deceptive perturbations, which are carefully crafted to induce specific misclassifications in the attacker's extracted model. By strategically injecting these perturbations, the defender can create "honeypot" regions in the input space that lead to predictable errors in stolen models, facilitating detection of unauthorized copies. These adversarial perturbation methods offer strong protection against extraction attacks but may require more computational resources and careful tuning to balance security and legitimate model performance.

\noindent\textbf{Noise Injection.} A more general approach to data perturbation involves injecting controlled noise into various stages of the model's operation. \citet{wu2024efficient} propose using a noise transition matrix to efficiently perturb the predicted posteriors of the target model. This method injects noise in a linear manner, allowing for seamless integration into existing systems with minimal overhead. The noise transition matrix is optimized through a bi-level optimization framework that balances model fidelity for legitimate users with adversarial protection against extraction attempts. Noise injection techniques offer flexibility in terms of where and how perturbations are applied, allowing defenders to tailor the approach to their specific use case and threat model. However, careful calibration is necessary to ensure that the injected noise effectively disrupts extraction attacks without overly degrading the model's utility for intended applications.

Data perturbation methods offer a diverse toolkit for defending against MEAs, each with distinct characteristics and trade-offs. Input perturbation techniques excel in scenarios where defenders can control query preprocessing and reliably distinguish between legitimate and potentially malicious inputs, offering a lightweight defense mechanism. However, they may struggle against sophisticated attackers who can generate queries that closely mimic legitimate input distributions. Output perturbation methods provide fine-grained control over the information revealed by model predictions, allowing for dynamic adaptation to different query types and attack patterns. These techniques can effectively obscure true decision boundaries but may introduce a delicate balance between security and utility for legitimate users. Adversarial perturbation approaches leverage techniques from adversarial example research to proactively alter model behavior in ways difficult for attackers to reverse-engineer. While offering strong protection, these methods often demand greater computational resources and expertise to implement effectively. Noise injection techniques, such as using noise transition matrices, strike a balance between simplicity and effectiveness. They offer flexibility in perturbation application and can be seamlessly integrated into existing systems, but require careful calibration to maintain model utility. The selection of a perturbation method hinges on factors including the specific threat model, computational constraints, and the desired security-utility trade-off. Input perturbation may be preferable when query preprocessing is already integrated into the pipeline, while output perturbation suits applications requiring precise control over model outputs. High-security environments might favor adversarial perturbation despite its computational overhead, whereas noise injection methods provide a versatile, plug-and-play solution adaptable to various deployment scenarios. It's crucial to recognize that while these methods significantly elevate defenses against extraction attacks, they are not infallible. Ongoing research is necessary to develop more adaptive, context-aware strategies that can dynamically adjust to evolving attack patterns while minimizing impact on legitimate users, as sophisticated attackers may devise methods to circumvent specific perturbation techniques.
\subsubsection{Model Modification}

Model modification techniques aim to protect deep neural networks (DNNs) from extraction attacks by altering the model architecture or parameters in ways that preserve functionality while obscuring the model's true structure. These approaches can be broadly categorized into three main strategies: architecture obfuscation, parameter obfuscation, and input transformation. The general formulation for model modification techniques can be expressed as
\begin{equation}
\mathcal{M}' = g(\mathcal{M}, K),
\end{equation}
where $\mathcal{M}$ is the original model, $K$ is a secret key or set of parameters, $g$ is a modification function, and $\mathcal{M}'$ is the protected model. The goal is to design $g$ such that $f_{\mathcal{M}'}(x) \approx f_{\mathcal{M}}(x)$ for legitimate inputs $x$, while making it difficult for an attacker to recover $\mathcal{M}$ from $\mathcal{M}'$ without knowledge of $K$.

\noindent\textbf{Architecture Obfuscation.} This approach focuses on modifying the structure of the neural network to make it more difficult for attackers to understand or replicate. \citet{szentannai2020preventing} proposed a method to prevent weight stealing by obfuscating the network structure. Their approach involves creating redundant paths and dummy neurons to confuse potential attackers while maintaining the original functionality. Similarly, \citet{li2021neurobfuscator} introduced NeurObfuscator, a full-stack obfuscation tool that applies various techniques such as layer splitting, neuron pruning, and activation function substitution to protect the model architecture. These methods effectively increase the search space for attackers attempting to reverse-engineer the model structure. However, they may introduce computational overhead and potentially impact model performance if not carefully implemented. A novel approach proposed by \citet{lin2020bident} utilizes a bident structure, where the network is split into two sub-networks deployed in separate environments (trusted and untrusted). This method not only protects the model structure but also leverages trusted execution environments for enhanced security.
A hybrid approach that combines multiple protection strategies is presented by \citet{xu2018deepobfuscation}. Their DeepObfuscation method secures the structure of convolutional neural networks via knowledge distillation, effectively combining architecture obfuscation with a form of model compression.
\citet{li2024translinkguard} proposed TransLinkGuard, a plug-and-play protection approach for edge-deployed transformer models. This method introduces a lightweight authorization module in a secure environment (e.g., TEE) to authorize each request based on its input, effectively creating a barrier against unauthorized access while addressing multiple protection properties simultaneously. TransLinkGuard is suited for scenarios where models are deployed on untrusted devices.

\noindent\textbf{Parameter Obfuscation.} This category of techniques focuses on modifying the model parameters to make them more resistant to extraction attempts. \citet{goldstein2021preventing} proposed a hardware-based approach to prevent DNN model IP theft by obfuscating the model parameters using a secret key. Their method involves embedding the key into the hardware implementation of the model, making it extremely difficult for attackers to extract the correct parameters without physical access to the device. Another interesting approach in this category is the work by \citet{sun2024streamlining}, which proposes a streamlined DNN obfuscation method to defend against model stealing attacks. Their technique focuses on selectively obfuscating critical parameters while minimizing the impact on model performance and inference speed. These parameter obfuscation methods provide strong protection against extraction attacks but may require specialized hardware or careful implementation to avoid significant performance degradation.
An alternative approach to parameter obfuscation was proposed by \citet{chabanne2020protection}, who introduced a method of adding parasitic layers to neural networks. These layers approximate a noisy identity mapping, keeping the underlying model's predictions mostly unchanged while complicating reverse-engineering attempts.

\noindent\textbf{Input Transformation.} This approach involves modifying the input data or the way the model processes inputs to protect against extraction attacks. \citet{olney2022protecting} introduced an architecture-agnostic input obfuscation method to protect DNN intellectual property. Their technique applies a transformation to the input data that can only be reversed with a secret key, effectively preventing unauthorized use of the model. Similarly, \citet{maungmaung2021protection} proposed a method to protect trained CNN models using a secret key for input transformation. These input transformation techniques have the advantage of being relatively easy to implement and can be applied to existing models without modifying their internal structure. However, they may introduce some computational overhead during inference and require secure key management.

Architecture obfuscation methods are particularly effective for protecting proprietary model designs but may introduce more significant overhead. Parameter obfuscation techniques provide strong protection for model weights and biases, especially when implemented in hardware, but may require specialized infrastructure. Input transformation approaches offer flexibility and ease of implementation but need careful key management. The choice of protection method depends on the specific threat model, deployment environment, and performance requirements. For cloud-based services, input transformation or software-based obfuscation might be more suitable, while hardware-based parameter obfuscation could be ideal for edge devices. Hybrid approaches that combine multiple techniques may offer the most robust protection but require careful design to balance security, performance, and usability. As the field evolves, we can expect to see more sophisticated obfuscation techniques that provide stronger protection with minimal impact on model functionality and efficiency.

\subsubsection{Access Control}
Access control mechanisms serve as a proactive defense against MEAs by restricting the adversary's ability to query the target model extensively. This approach can be formulated as
\begin{equation}
\mathcal{D}(M, Q, B) =
\begin{cases}
M(Q) & \text{if } |Q| \leq B \\
\text{Reject} & \text{otherwise}
\end{cases},
\end{equation}
where $M$ is the target model, $Q$ is the set of queries, $B$ is the query budget, and $\mathcal{D}$ is the defense.

\citet{dziedzic} proposed a novel defense strategy that employs a calibrated proof-of-work mechanism to increase the computational cost of MEAs. Their method requires users to complete a computational task before accessing the model's predictions, effectively deterring large-scale querying without significantly impacting legitimate users. This approach is particularly effective against attacks that rely on massive query volumes. In contrast, \citet{kesarwani2018model} introduce a model extraction warning system for MLaaS platforms. Their system monitors API query patterns to detect potential extraction attempts, allowing service providers to take preventive actions. This method focuses on early detection rather than imposing direct query limitations.  

Access control techniques offer a straightforward and effective first line of defense against MEAs. These methods are particularly suitable for cloud-based ML services where controlling API access is feasible. The main advantage of these approaches is their simplicity and low computational overhead. However, they face challenges in distinguishing between legitimate intensive use and malicious extraction attempts, potentially impacting service usability. The proof-of-work approach offers a more flexible solution by allowing high-volume querying at an increased computational cost, but it may still inconvenience some legitimate users. These methods are most effective when combined with other defense strategies, such as output perturbation or watermarking.
\subsubsection{Adversarial Training}
Adversarial training approaches leverage the principles of adversarial machine learning to defend against MEAs. The general formulation for these methods can be expressed as
\begin{equation}
\min_{\theta} \mathbb{E}_{(x,y) \sim \mathcal{D}} [\max{\delta \in \Delta} \mathcal{L}(f_\theta(x + \delta), y)],
\end{equation}
where $f_\theta$ is the target model with parameters $\theta$, $\mathcal{D}$ is the data distribution, $\Delta$ is the set of allowable perturbations, and $\mathcal{L}$ is the loss function.

~\citet{yilmaz} introduce the Adversarial Sparse Teacher (AST) method, which trains a teacher model using adversarial examples to produce sparse logit responses and increase the entropy of the output distribution. This approach embeds altered logits into the output while keeping the primary response slightly higher, effectively confusing potential attackers. \citet{zhang2021seat} propose SEAT, a similarity encoder trained by adversarial training to detect malicious queries indicative of MEAs. SEAT leverages the insight that extraction queries often differ from benign ones in their similarity structure. \citet{chen2021ast} present DAS-AST, which employs an adaptive softmax transformation to defend against model stealing. This method dynamically adjusts the model's output based on detected extraction attempts, making it more challenging for attackers to accurately replicate the model's behavior.

Adversarial training techniques offer a proactive defense against model extraction by fundamentally altering the behavior of the target model in ways that are difficult for attackers to exploit. These methods are particularly effective against query-based attacks, as they can significantly degrade the quality of information obtained through malicious queries. The main advantage of adversarial training approaches is their ability to maintain high performance for legitimate users while substantially increasing the difficulty and cost of extraction attempts. However, they may introduce computational overhead during both training and inference, potentially impacting model efficiency. Additionally, careful calibration is necessary to balance defense strength with model utility. These techniques are most suitable for high-stakes applications where the cost of model theft significantly outweighs the additional computational requirements. They can be particularly effective when combined with detection methods, allowing for adaptive defense strategies that apply more aggressive protections only when an attack is suspected.

\subsection{Compositional Defenses}

\subsubsection{Holistic Defense Strategies}

Holistic defense strategies against MEAs aim to provide comprehensive protection by addressing multiple vulnerabilities simultaneously. These approaches can be broadly categorized into two main types: \textit{optimization-based defenses and ensemble methods}.

\noindent\textbf{Optimization-based Defenses.} Several approaches leverage optimization techniques to create robust defenses. \cite{zhang2024defense,mori2021bodame,xian2022framework,jiang2023comprehensive} propose methods that formulate the defense problem as complex optimization tasks. These methods aim to find an optimal balance between model utility and protection against extraction attacks.
Gradient-based Ranking Optimization (GRO) \cite{zhang2024defense} and BODAME \cite{mori2021bodame} use bilevel optimization to create substitute models that diverge from potential attackers' models. These approaches are particularly effective against attacks that rely on querying the model with out-of-distribution samples. However, they may introduce computational overhead during the training phase and require careful tuning to maintain model performance on legitimate queries.
The framework proposed by \cite{xian2022framework} provides a theoretical foundation for understanding the trade-offs between model utility and privacy. This approach offers insights into optimal strategies for both attackers and defenders, potentially guiding the development of more effective defenses. However, translating these theoretical insights into practical defenses remains a challenge.
\citet{jiang2023comprehensive} proposed a comprehensive defense framework that combines multiple techniques, including optimization-based methods. This approach aims to provide broader protection against various types of extraction attacks by integrating different defense strategies. While this method offers increased robustness, it may also be more complex to implement and fine-tune for specific scenarios.

\noindent\textbf{Ensemble Methods.} Ensemble methods leverage the power of multiple models or architectural modifications to create a more robust defense. \cite{kariyappa2021protecting,li2024translinkguard,chabanne2020protection} propose various techniques that combine model diversity with structural changes to enhance protection.
The Ensemble of Diverse Models (EDM) \cite{kariyappa2021protecting} trains multiple models to produce dissimilar predictions for out-of-distribution inputs, creating discontinuities that hinder extraction attempts. This method is particularly effective against attacks that use synthetic or out-of-distribution queries. However, managing multiple models can increase computational and storage requirements.
TransLinkGuard \cite{li2024translinkguard} introduces a secure authorization module for edge-deployed models, effectively creating a barrier against unauthorized access. This approach combines architectural modification with a form of ensemble decision-making, as it uses different components for authorization and prediction. It is particularly suited for scenarios where models are deployed on potentially untrusted devices, although it may introduce latency in model inference and requires secure hardware support.
\citet{chabanne2020protection} proposed adding parasitic layers to neural networks, which can be seen as a form of architectural ensemble. These layers approximate a noisy identity mapping, keeping the underlying model's predictions mostly unchanged while complicating reverse-engineering attempts. This technique offers a balance between protection and maintaining model performance, although its effectiveness may vary depending on the specific attack scenario.

\subsubsection{Specialized Protection Mechanisms}

Recent research has developed various specialized defense mechanisms against MEAs, which can be broadly categorized into three main approaches: response manipulation, structural modifications, and access control mechanisms.

\noindent\textbf{Response Manipulation.} A significant portion of defenses focuses on manipulating model responses to thwart extraction attempts while maintaining utility for legitimate users. \cite{yilmaz, wu2024efficient,kariyappa2020defending,orekondy2019prediction} propose methods that modify model outputs through adversarial perturbations or noise injection. The Adversarial Sparse Teacher \cite{yilmaz} generates sparse logit responses with increased entropy using adversarial examples, while maintaining prediction accuracy. Similarly, noise transition matrices \cite{wu2024efficient} introduce carefully calibrated perturbations to predicted posteriors. These approaches are particularly effective against query-based attacks but may suffer from reduced effectiveness against adaptive adversaries who can average multiple queries.
Another line of work exploits out-of-distribution (OOD) behaviors and decision boundaries. \cite{liang2024defending,xie2024same,lee2022model} leverage the distinct characteristics of OOD queries typically used in extraction attacks. Decision boundary confusion and feature learning techniques \cite{liang2024defending} specifically target the boundary regions where attackers often probe, while sample reconstruction methods \cite{xie2024same} aim to detect and mitigate extraction attempts based on query patterns. These defenses show strong performance against black-box attacks but may be vulnerable to attacks using in-distribution queries.

\noindent\textbf{Structural Protection.} Several approaches modify model architectures or integrate protective components directly into the network structure. \cite{lin2020bident,guo2023isolation,maungmaung2021protection} propose architectural modifications that inherently resist extraction. The bident structure \cite{lin2020bident} splits the network into two parts with one running in a secure environment, while isolation and induction techniques \cite{guo2023isolation} create model components specifically designed to mislead extraction attempts. These methods offer strong protection against white-box attacks but may introduce computational overhead.
Verification-based approaches \cite{li2022defending,zhang2023categorical,dubinski2023bucks} embed verifiable features or signatures into the model. External feature verification \cite{li2022defending} and categorical inference poisoning \cite{zhang2023categorical} provide mechanisms to detect unauthorized model copies. These methods offer robust protection against model replication but require careful design to avoid impacting legitimate use cases.

\noindent\textbf{Access Control and Cost Imposition.} A third category focuses on controlling model access and increasing extraction costs. \cite{dziedzic,mazeika2022steer} implement mechanisms that make extraction attacks prohibitively expensive or time-consuming. Proof-of-work schemes \cite{dziedzic} impose computational costs on queries, while gradient redirection \cite{mazeika2022steer} strategically guides potential attackers away from sensitive model components. \cite{chen2021ast,zheng2019bdpl} propose adaptive access control mechanisms that dynamically adjust model responses based on user behavior patterns.
The Bucks for Buckets approach \cite{dubinski2023bucks} introduces an active defense system that monitors embedding space coverage to identify potential extraction attempts. Information laundering techniques \cite{wang2020information} provide an additional layer of protection by obfuscating model responses while preserving their utility.

These specialized defenses demonstrate several key trade-offs. Response manipulation methods offer flexibility and ease of implementation but may be vulnerable to adaptive attacks. Structural protections provide stronger guarantees but often at the cost of increased complexity or computational overhead. Access control mechanisms offer practical deployment options but require careful balancing of security and usability. The effectiveness of these approaches largely depends on the specific attack scenario and deployment constraints, suggesting that combining multiple defense strategies may be necessary for comprehensive protection.

\subsection{Defenses Tailored for Specific Data Modalities}

\subsubsection{Defenses on Text Models}
Defending text models against extraction attacks presents unique challenges due to the complex nature of language and diverse attack vulnerabilities. Recent research has focused on developing innovative defense mechanisms tailored to text-based models. \citet{he2022cater} introduced CATER, a conditional watermarking framework to protect text generation APIs from imitation attacks. CATER optimizes watermarking rules to minimize distortion of word distributions while maximizing changes in conditional word selections. Patil et al. \cite{patil2023can} explored deleting sensitive information directly from LLM weights, proposing methods like Head Projection Defense and Max-Entropy Defense to remove sensitive information from both final outputs and intermediate hidden states. Addressing edge-deployed models, \citet{li2024translinkguard} proposed TransLinkGuard, a plug-and-play protection approach utilizing a lightweight authorization module in a secure environment to safeguard against model stealing attacks while maintaining functionality.
Text data's sequential and contextual nature presents unique challenges for defending language models against extraction attacks. The high dimensionality and semantic complexity of textual data create diverse attack vulnerabilities, making it difficult to implement comprehensive defenses. Watermarking techniques for text must preserve linguistic coherence and semantic meaning while embedding protective features, a delicate balance that is more challenging than in other data modalities. Deletion of sensitive information from model weights is complicated by the distributed nature of knowledge representation in language models, where concepts are often encoded across multiple layers and neurons. The contextual nature of language also makes it challenging to identify and remove specific information without affecting related concepts or general understanding.
\subsubsection{Defenses on Vision Models}
Defending vision models against extraction attacks has become increasingly critical as these models find widespread use in various applications. Several key defense strategies have emerged, each addressing different aspects of model protection. Watermarking techniques have been adapted to protect vision models, as demonstrated by DeepiSign \cite{abuadbba2021deepisign} and the work by \citet{tang2023exposing}. These approaches embed invisible, fragile watermarks into CNN models to protect their integrity and authenticity. Adaptive misinformation, introduced by \citet{kariyappa2020defending}, trains the defender's model to produce high-confidence predictions on in-distribution samples while giving random predictions on out-of-distribution samples. \citet{maungmaung2021protection} proposed a method to protect trained CNN models using a secret key, preprocessing input images with block-wise transformations. \citet{wu2024efficient} developed an efficient model stealing defense using a noise transition matrix, injecting noise into predicted posteriors in a linear manner.
The defense of vision models faces unique challenges due to the high-dimensional and spatially structured nature of image data. This characteristic makes models vulnerable to attacks exploiting both local features and global patterns. Watermarking techniques must balance robustness against common image transformations with imperceptibility to human observers. Adaptive defenses like misinformation and noise injection strategies struggle to maintain a balance between model protection and performance on legitimate inputs.
\subsubsection{Defenses on Graph Models}
Defenses for GNNs have emerged to counter various attack vectors, broadly categorized into watermarking techniques and privacy-preserving methods. Watermarking approaches aim to protect intellectual property rights of GNNs. \citet{zhao2021watermarking} proposed using random graphs as triggers to embed watermarks, allowing ownership verification without compromising model performance. \citet{waheed2023grove} introduced GrOVe, an ownership verification scheme using embeddings, effective even when suspect models use the same training data and architecture. \citet{you2024gnnguard} developed GNNGuard, a fingerprinting framework demonstrating effectiveness across various GNN architectures and tasks. For link prediction specifically, \citet{bachina2024genie} proposed GENIE, using a novel backdoor attack to create trigger sets for watermarking. Advancing this approach, \citet{dai2024pregip} introduced PreGIP, a framework for watermarking pretrained GNN encoders, incorporating task-free watermarking loss and finetuning-resistant watermark injection. Beyond ownership verification, \citet{cheng2025atom} detect query-based extraction on GNN services in real time by modeling sequential, structure-aware query patterns. On the privacy-preserving front, \citet{sajadmanesh} proposed GAP, a differentially private GNN using aggregation perturbation, providing both edge-level and node-level privacy guarantees while allowing multi-hop neighborhood aggregations. \citet{pei2024privacy} developed a privacy-enhanced GNN for decentralized local graphs, further addressing privacy concerns in distributed settings. These defense mechanisms address unique challenges in the graph domain: preserving complex structural information while protecting against attacks, adapting to various GNN architectures and tasks, and balancing privacy guarantees with the ability to leverage multi-hop information. The interconnected nature of graph data presents a particular challenge, as protecting one aspect (e.g., node features) may inadvertently expose vulnerabilities in another (e.g., edge information). As the field evolves, future research may need to focus on developing more robust watermarking techniques and privacy-preserving methods that can protect against increasingly attacks targeting the interdependence between node features and graph structure.

\section{MEAs Research in Different Computing Environments} \label{environment}
\noindent\textbf{Intuition.} Our analysis of model extraction across different computing environments reveals how the unique characteristics of each deployment scenario shape both attack vectors and defense strategies. We examine three fundamental computing paradigms that present distinct security challenges and opportunities. Cloud computing, with its service-oriented architecture, must balance accessibility with security, leading to unique attack scenarios through API interfaces and corresponding defense mechanisms like query monitoring and access control. Edge computing introduces challenges stemming from physical device accessibility and resource constraints, where attackers can exploit hardware side-channels while defenders must optimize protection mechanisms within limited computational budgets. Federated learning presents perhaps the most complex scenario, where the collaborative nature of model training creates novel attack vulnerabilities through gradient sharing, requiring sophisticated defense strategies that preserve both privacy and training effectiveness. This systematic examination demonstrates how environmental constraints fundamentally influence both attack methodologies and defense designs: cloud environments must protect against API-based attacks while maintaining service quality, edge devices require lightweight yet robust defenses against physical attacks, and federated learning needs to balance collaborative benefits with gradient leakage protection. By analyzing these environment-specific challenges, we provide insights into how the deployment context fundamentally shapes security considerations in machine learning systems, guiding the development of targeted protection mechanisms that address the unique vulnerabilities of each computing paradigm.
\begin{figure}[htbp]
    \centering
    \includegraphics[width=0.98\textwidth]{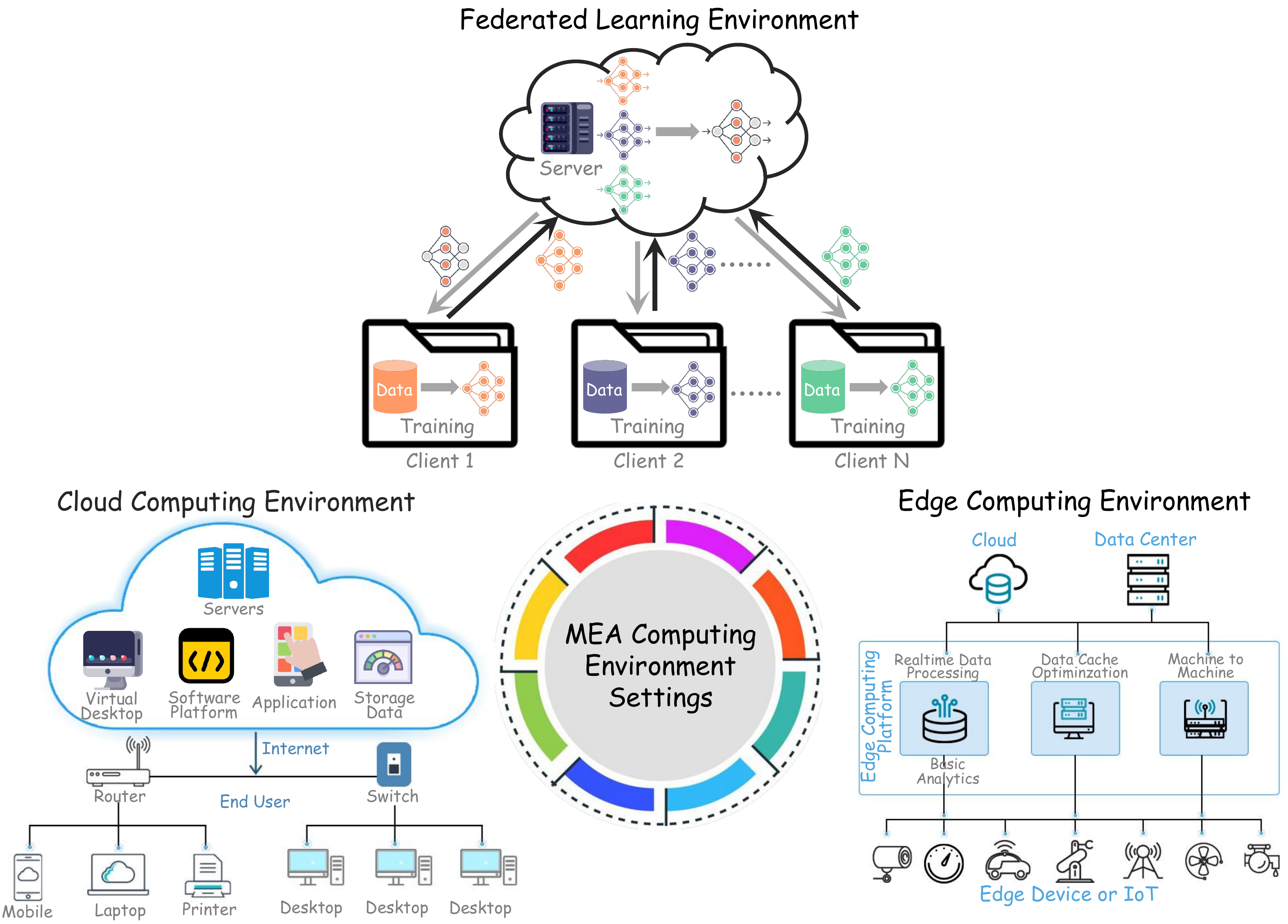}
    \caption{Detailed illustration of model extraction attack under different computing environments.}
    \label{fig:compute_environment}
\end{figure}

\subsection{Cloud Computing Environment}
\textbf{Cloud Computing.} Cloud computing\cite{qian2009cloud, antonopoulos2010cloud} is a computing paradigm that provides on-demand access to a shared pool of configurable computing resources\cite{baliga2010green,exposito2013general,azodolmolky2013cloud,ali2013cloud} (eg., networks, servers, storage, GPUs, applications) over the internet. This model enables ubiquitous, convenient, and on-demand network access to these resources, which can be rapidly provisioned and released with minimal management effort or service provider interaction. Cloud computing typically operates on a pay-as-you-go model\cite{Abbasov2014cloud}, allowing users to scale resources up or down based on their demands.

\noindent\textbf{MEAs in MLaaS Platforms.}
In the context of MLaaS platforms, MEAs present unique challenges due to the limited API access, query rate limitations, and black-box nature of the deployed models \cite{kesarwani2018model, gong2020model}. Adversaries aim to minimize the number of queries required to extract the target model while operating within the constraints imposed by the cloud platform \cite{yang2024swifttheft}. This necessitates the development of query-efficient attack strategies that can effectively steal the model's functionality based solely on its observable input-output behavior, without knowledge of the model's internal architecture or parameters \cite{gong2020model}.

\noindent\textbf{Defense Strategies and Challenges.}
To counter model extraction threats, cloud service providers must implement robust defense mechanisms such as query filtering, rate limiting, and API access control \cite{kesarwani2018model}. Additionally, techniques like model watermarking and ownership verification can help identify stolen models and prove intellectual property rights \cite{yang2024swifttheft, zhao2025design}. Beyond OOD-based detection, \citet{cheng2025misleader} propose MISLEADER, an active defense for MLaaS that trains an ensemble of distilled models with a bi-level objective and augmentation to preserve utility while reducing extractability. However, implementing these defenses in a cloud environment presents challenges in balancing security with usability and performance. Overly restrictive measures may hinder legitimate users and impact service quality. Therefore, cloud providers must carefully navigate these trade-offs and develop effective defense strategies that protect their models without compromising the user experience. Continued research and innovation are crucial to address the evolving landscape of model extraction threats in the cloud computing paradigm.

\subsection{Edge Computing Environment}
\textbf{Edge Computing.} Edge computing\cite{khan2019edge,cao2020overview} is an emerging paradigm that brings computation and data storage closer to the location where it is needed, to improve response times and save bandwidth \cite{shi2016edge}. Unlike cloud computing where data processing occurs in centralized data centers, edge computing pushes the processing to the network edge, near the data sources \cite{satyanarayanan2017emergence}. The main characteristics of edge computing include proximity, dense geographical distribution, support for mobility, location awareness, low latency, and heterogeneity \cite{bonomi2012fog,ahmed2017role,manzalini2018edge,taleb2017multi,khan2019edge,mao2017survey}. Edge computing complements cloud computing by addressing issues like high latency, limited bandwidth, and privacy concerns associated with sending all data to centralized cloud servers \cite{garcia2015edge,wu2023fedcache,wu2023knowledge}. It enables a new breed of applications and services, particularly in areas like Internet of Things (IoT), autonomous vehicles, and augmented reality \cite{chiang2016fog}.

\noindent\textbf{MEAs in Edge Computing.}
The unique characteristics of edge computing environments introduce distinct settings and challenges\cite{varghese2016challenges} for MEAs. Edge devices, such as smartphones, IoT devices, and embedded systems, are typically resource-constrained in terms of memory, computation power, and energy \cite{kumar2021resource,rakin2022deepsteal}. This resource scarcity necessitates the use of compact and efficient ML models, which may be more susceptible to extraction attacks \cite{ren2024demistify,nazari2024llm}. Moreover, the physical proximity and accessibility of edge devices to potential attackers increase the risk of side-channel attacks that exploit hardware vulnerabilities to extract model information \cite{nazari2024llm,batina2019csi,yu2020deepem}.
The distributed nature of edge computing also presents unique challenges for MEAs. Edge devices often collaborate and communicate with each other to perform distributed inference or federated learning \cite{wu2023fedcache,wu2023knowledge}. This distributed setting expands the attack vulnerability, as an adversary can target multiple devices to piece together the complete model \cite{meyers2024trained,nasr2019comprehensive}. From a technical perspective, MEAs in edge computing leverage various side-channels, such as power consumption \cite{breier2022sniff,liu2024model}, electromagnetic emanations \cite{batina2019csi,yu2020deepem}, timing information \cite{hu2019neural,Yan2020cache}, and memory access patterns \cite{rakin2022deepsteal, Hu2020DeepSniffer}. These side-channels can be exploited to infer the architecture, parameters, and even the training data of the target models.

\noindent\textbf{Defenses against MEAs in Edge Computing.}
Defending against MEAs in edge computing environments is a challenging task due to the resource constraints and heterogeneity of edge devices. The design of robust and generalizable defenses is complicated by the variety of hardware, software, and data distributions across different edge devices \cite{sun2024layer,hu2019neural}. Defenses against MEAs involve techniques such as side-channel masking, secure enclaves, oblivious computation, and hardware-based security primitives \cite{volos2018graviton,tramer2018slalom,van2018foreshadow}.
Side-channel masking techniques aim to obscure the relationship between the sensitive model information and the observable side-channel signals. This can be achieved through techniques such as power balancing, noise injection, and randomization of the computation \cite{standaert2010introduction}. Secure enclaves, such as Intel SGX and ARM TrustZone, provide isolated execution environments that protect the confidentiality and integrity of the model computation \cite{volos2018graviton,lee2019occlumency}. Oblivious computation techniques, such as homomorphic encryption and secure multi-party computation, enable the execution of ML models without revealing the model parameters or the input data \cite{gilad2016cryptonets,mohassel2017secureml}. Hardware-based security primitives, such as physically unclonable functions (PUFs) and true random number generators (TRNGs), can be used to generate device-specific keys and to ensure the freshness and randomness of the cryptographic operations \cite{delvaux2017security,herder2014physical}.The resource constraints, physical accessibility, and distributed nature of edge computing environments present unique challenges and opportunities for both MEAs and defenses. The interplay between the hardware and software stack, as well as the trade-offs between performance, privacy, and security, require careful consideration when designing and deploying ML models on edge devices. Addressing these challenges calls for a multidisciplinary approach that spans machine learning, hardware security, cryptography, and systems design.

\subsection{Federated Learning Environment}
\textbf{Federated Learning.} Federated Learning (FL) is a machine learning paradigm that enables training models on distributed datasets without sharing raw data \cite{mcmahan2017communication}. In FL, multiple clients collaboratively train a model under the orchestration of a central server, while keeping their data locally. The general formulation of FL can be expressed as
\begin{equation}
\min_{\theta} \sum_{k=1}^{K} w_k L_k(\mathcal{S}_k, \theta),
\end{equation}
where $\theta$ represents the model parameters, $K$ is the number of clients, $w_k$ is the weight of client $k$, $L_k$ is the local loss function, and $\mathcal{S}_k$ is the local dataset of client $k$.

\noindent\textbf{MEAs in Federated Learning.}
Federated learning (FL) has emerged as a promising approach for collaborative learning across multiple parties while preserving data privacy. However, the distributed nature of FL introduces unique vulnerabilities and challenges for MEAs. In FL, the global model is updated by aggregating the local model updates from multiple participants, which can potentially leak sensitive information about the participants' local data and models \cite{nasr2019comprehensive}.
MEAs in FL can be broadly categorized into two types of attacks: gradient leakage attacks and model update analysis. Gradient leakage attacks aim to infer sensitive information about the participants' local data or models by analyzing the gradients exchanged during the FL process \cite{zhu2019deep,zhao2020idlg}. Model update analysis attacks focus on inferring the local model architectures, parameters, or hyperparameters by analyzing the model updates shared by the participants \cite{wang2019beyond,ganju2018property}. These attacks exploit the fact that the model updates contain information about the local models and can be used to reconstruct or infer the local models.
The decentralized nature of FL poses unique challenges for defending against MEAs. In FL, there is no central authority that can enforce security policies or monitor the participants' behavior. Moreover, the participants may have different levels of trust and may not be willing to share their local data or models. Therefore, defending against MEAs in FL requires the development of privacy-preserving protocols that can protect the confidentiality of the participants' local data and models while enabling collaborative learning \cite{bonawitz2017practical,qi2023differentially,olatunji2023private}.

\noindent\textbf{Defenses against MEAs in Federated Learning.}
Defending against MEAs in FL requires a combination of cryptographic techniques, secure aggregation protocols, and privacy-preserving mechanisms. One common approach is to use secure multiparty computation (SMC) protocols, such as homomorphic encryption and secret sharing, to enable the aggregation of the local model updates without revealing the individual updates \cite{mohassel2017secureml,zhang2020batchcrypt}. Another approach is to use differential privacy techniques to add noise to the model updates before sharing them, which can limit the amount of information that can be inferred from the updates \cite{abadi2016deep,geyer2017differentially}.
In addition to cryptographic techniques, FL systems can also employ secure aggregation protocols that ensure the confidentiality and integrity of the aggregated model updates. For example, the secure aggregation protocol proposed by \cite{bonawitz2017practical} uses a combination of secret sharing and homomorphic encryption to enable the secure aggregation of the local model updates without revealing the individual updates. Other secure aggregation protocols, such as \cite{qin2020selective,chen2021communication}, use techniques such as masked model updates and random perturbations to protect the privacy of the participants' local models.
However, the development of effective defenses against MEAs in FL is challenging due to the distributed nature of FL and the trade-offs between privacy, utility, and efficiency. Designing privacy-preserving FL protocols that can provide strong privacy guarantees while maintaining the accuracy and convergence of the global model is an active area of research \cite{yang2019federated,li2020federated}. Moreover, the heterogeneity of the participants' local data and models poses challenges for the development of generalizable and scalable defense mechanisms \cite{li2019convergence,briggs2020federated}.
MEAs in FL exploits the vulnerabilities introduced by the distributed nature of FL and the sharing of model updates among the participants. Defending against these attacks requires the development of privacy-preserving protocols and secure aggregation mechanisms that can protect the confidentiality of the participants' local data and models while enabling collaborative learning. The development of effective defenses calls for a multidisciplinary approach that combines cryptography, privacy-preserving machine learning, and distributed systems.

\section{Fundamental Evaluation Protocol} \label{evaluation}
When evaluating the performance of MEAs and their corresponding defense measures, researchers typically employ a range of common machine learning evaluation metrics. These metrics provide a framework for assessing model performance and the efficacy of both attacks and defenses.

\noindent\textbf{Accuracy.} This metric quantifies the proportion of correct predictions made by a model across all instances in the dataset. It is particularly useful for balanced datasets and provides an overall measure of model performance.

\noindent\textbf{Area Under the Curve (AUC).} AUC represents the model's ability to discriminate between classes. It is especially valuable for imbalanced datasets and provides a single scalar value that summarizes the model's performance across all possible classification thresholds.

\noindent\textbf{F1 Score.} As the harmonic mean of precision and recall, the F1 score offers a balanced measure of a model's performance, particularly useful when dealing with imbalanced datasets.

\noindent\textbf{Precision and Recall.} These complementary metrics provide insights into a model's ability to avoid false positives (precision) and its capability to identify all positive instances (recall). They are crucial for understanding the trade-offs in model performance.

Although these common metrics provide a solid foundation for evaluation, the unique characteristics of MEAs require the development of specialized metrics. These MEAs-specific metrics are designed to capture the nuanced aspects of attack effectiveness and defense efficacy. In the following subsections, we will delve into these specialized metrics, elucidating their formulations and significance in the context of MEAs and defenses.

\subsection{Metrics for Evaluating Model Extraction Attacks}
In this subsection, we present detailed discussions of commonly used metrics for evaluating the performance of MEAs, considering different mainstream settings and attack types identified in our taxonomy. Given the diversity of model extraction methodologies, evaluation protocols differ significantly between various attack scenarios.


\noindent\textbf{Extraction Accuracy} \cite{tramer2016stealing,orekondy2019knockoff,jagielski2020high}. Extraction accuracy is widely adopted for evaluating query-based and data-driven attacks, reflecting the ability of the extracted model to accurately mimic the predictions of the target model in a given test dataset. Higher accuracy indicates a more successful extraction, demonstrating that the attacker has effectively captured the target model's input-output mapping, thus compromising its intellectual property.

\noindent\textbf{Fidelity} \cite{jagielski2020high,krishna2019thieves,pal2020activethief}. Fidelity measures the consistency between the extracted model's predictions and those of the target model across identical test samples. Particularly relevant for substitute model training and meta-model attacks, fidelity captures how precisely the attacker has replicated the decision boundaries of the target model. A higher fidelity value implies that the extracted model closely mirrors the decision-making process of the original model, making it highly valuable for downstream adversarial applications.

\noindent\textbf{Extraction Efficiency} \cite{pal2020activethief,yu2020cloudleak}. Extraction efficiency specifically evaluates attacks that face query or computational resource constraints. Quantifies the relationship between the extraction accuracy achieved and the total number of queries made to the target model. High extraction efficiency means that the attacker can achieve substantial extraction accuracy with relatively fewer queries, indicating a potent threat under limited query budgets or computational resources.

\noindent\textbf{Transferability} \cite{orekondy2019knockoff,krishna2019thieves}. Transferability measures the ability of an extracted model to generalize effectively to different but related tasks or datasets, beyond the original context in which the target model was trained. This metric is particularly crucial for attacks on large-scale language models, vision models, and graph models, reflecting the potential for broader misuse of extracted models in various downstream tasks or further adversarial contexts.

\noindent\textbf{Model Parameter Similarity} \cite{jagielski2020high,carlini2021extracting}. Model Parameter Similarity evaluates the exactness with which the attacker can reconstruct the original parameters of the target model, including weights and biases. Primarily used in equation-solving and recovery attacks, a high similarity indicates that the attacker has successfully obtained precise internal model parameters, posing a severe security threat, especially to proprietary and sensitive models.

\subsection{Metrics for Evaluating Defenses against Model Extraction Attacks}
In this subsection, we elaborate on metrics widely utilized for assessing the effectiveness of various defense mechanisms against MEAs, aligning closely with our taxonomy.

\noindent\textbf{Defense Success Rate} \cite{juuti2019prada,kariyappa2020defending,maini2021dataset}. Defense success rate broadly evaluates the effectiveness of defense mechanisms by measuring the frequency of successful defense against extraction attempts. Caculate the proportion of attacks prevented from reaching a predefined threshold of extraction accuracy or fidelity. Higher values are indicative of stronger and more reliable defense capabilities across different types of attacks.

\noindent\textbf{Utility-Security Trade-off} \cite{chandrasekaran2020exploring,kariyappa2020defending,maini2021dataset}. Utility-security trade-off is crucial for evaluating the effectiveness of defenses that inherently impact model utility, such as data perturbation, model modification, and differential privacy. This metric assesses the balance between maintaining the original predictive accuracy of the target model and enhancing its security against extraction attacks. A better trade-off indicates that the defense mechanism effectively protects the model without significantly compromising its original functionality or performance.

\noindent\textbf{Query Detectability} \cite{juuti2019prada,atli2020extraction,kariyappa2020defending}. Query detectability specifically measures how effectively a defense mechanism identifies malicious query patterns indicative of MEAs. Primarily utilized by monitoring-based defense systems, query detectability is commonly assessed through precision, recall, and F1 score metrics of attack detection classifiers. High scores imply that the defense is proficient in accurately and promptly detecting extraction attempts, thus facilitating timely defensive responses.

\noindent\textbf{Robustness to Adaptive Attacks} \cite{orekondy2019knockoff,kariyappa2020defending,maini2021dataset}. Robustness to Adaptive Attacks evaluates the resilience of defense mechanisms against adversaries who specifically adapt their attack strategies to circumvent known defenses. This metric assesses whether defenses remain effective even in highly strategic and informed adversarial scenarios. Higher robustness indicates that the defense mechanism can withstand evolving and sophisticated attack methodologies, representing a robust and comprehensive defensive posture.

\section{Real-world Application Scenarios} \label{app}
\noindent\textbf{Intuition.} In this section, we introduce several high-stake decision-making scenarios where MEAs pose significant risks. These applications demonstrate the critical nature of protecting machine learning models in various domains, highlighting the potential consequences of successful attacks and the unique challenges in implementing effective defenses. We focus on four key areas: finance, healthcare, autonomous vehicles, and cybersecurity, each with distinct vulnerabilities.

\noindent \textbf{Finance.} In the financial sector, MEAs pose significant risks to proprietary algorithms and sensitive data. Query-based attacks, such as those that employing substitute model training \cite{tramer2016stealing}, are particularly concerning due to the high-stakes nature of financial predictions. However, these attacks often require a large number of queries, which can be detected by monitoring-based defense methods \cite{juuti2019prada}. The challenge lies in distinguishing between legitimate high-frequency trading activities and malicious extraction attempts. Gradient-based attacks \cite{jagielski2020high} on models used for credit scoring or fraud detection could potentially reveal sensitive information about the underlying data. Defending against such attacks while maintaining model utility is challenging, as techniques like differential privacy \cite{abadi2016deep} may impact the accuracy of financial predictions. For example, a credit scoring model protected with too much noise might misclassify borderline applicants, leading to significant real-world consequences.

\noindent \textbf{Healthcare.} Healthcare models are particularly vulnerable to side-channel attacks \cite{batina2019csi} due to the sensitive nature of medical data and the potential for models to be deployed on various devices. These attacks could potentially reveal patient information or proprietary diagnostic techniques. Defending against such attacks often requires hardware-level protections \cite{goldstein2021preventing}, which can be costly to implement across various healthcare infrastructures. Explanation-based attacks \cite{milli2019model} pose a unique threat in this domain, as the interpretability of medical models is often crucial for clinical decision making. Balancing the explainability of the model with robustness against extraction is a significant challenge. For example, a model diagnosing rare diseases might be vulnerable to extraction through carefully crafted queries that exploit its explanation mechanisms, potentially compromising both patient privacy and the intellectual property of the healthcare provider.

\noindent \textbf{Autonomous Vehicles.} In the field of autonomous vehicles, MEAs could have severe safety implications. Vision models used for object detection and classification are particularly vulnerable to data-driven attacks \cite{orekondy2019knockoff}, which could potentially allow attackers to create adversarial scenarios. Defending against these attacks while maintaining real-time performance is challenging, as techniques such as watermarking \cite{adi2018turning} or model obfuscation \cite{zhou2022model, zhou2023nnsplitter, goldstein2021preventing, sun2024streamlining} can introduce latency that is unacceptable in safety-critical systems. Moreover, the distributed nature of autonomous vehicle systems, often involving edge computing \cite{mao2017survey}. For example, an attacker might attempt to extract the lane detection model from a vehicle's onboard computer, potentially allowing them to create scenarios that confuse the vehicle's navigation system.

\noindent \textbf{Cybersecurity.} Cybersecurity models, such as those used for intrusion detection or malware classification, are prime targets for MEAs. These models often need to process a wide range of inputs, making them vulnerable to query-based attacks that exploit this diversity \cite{chen2017targeted}. Defending against these attacks is particularly challenging due to the adversarial nature of the cybersecurity domain, where attackers are constantly evolving their techniques. Privacy-preserving machine learning techniques \cite{al2019privacy} offer some protection, but may struggle to keep up with the rapidly changing threat landscape. For example, a model used to detect new strains of malware might be vulnerable to extraction attacks that systematically probe its decision boundaries, potentially allowing attackers to create malware that evades detection.

\section{Future Directions and Challenges} \label{future}

\noindent\textbf{Attack Perspectives.} Model extraction attacks continue to evolve and present new challenges across different computing environments. Future research in advanced adaptive attacks should focus on developing more sophisticated techniques that can dynamically adjust strategies based on the target model's defensive responses, circumvent multiple layered defenses simultaneously, and maintain effectiveness while minimizing detection signatures. These adaptive attacks must be carefully designed to provide meaningful evaluation of defense mechanisms, rather than simply bypassing specific protective measures. In terms of environment-specific evolution, cloud computing attacks need to develop more efficient query optimization techniques that can overcome rate limiting while maintaining extraction effectiveness. Edge computing attacks should focus on integrating side-channel information with query-based approaches to enhance extraction accuracy while accounting for resource constraints. In federated learning environments, advanced gradient manipulation methods must be developed to extract information while evading increasingly sophisticated privacy-preserving mechanisms.

\noindent\textbf{Defense Perspectives.} Current defense mechanisms largely rely on empirical evaluation against specific attacks, creating a pressing need for more robust security guarantees. Future research should prioritize developing formal security guarantees for defense mechanisms, establishing certifiable bounds on model extraction success probability, and creating theoretical frameworks for measuring defense effectiveness. These certified guarantees would provide stronger assurances about defense robustness against any extraction attack within defined constraints. Environment-specific defenses present unique challenges: cloud environments require adaptive mechanisms that balance protection with service quality and latency requirements; edge computing needs lightweight defense methods suitable for resource-constrained devices; and federated learning demands privacy-preserving techniques that maintain collaborative learning benefits while preventing gradient leakage. Defense mechanisms must evolve to address these environment-specific challenges while maintaining practical deployability.

\noindent\textbf{Evaluation and Standardization.} The field currently lacks standardized evaluation frameworks that can comprehensively assess both attacks and defenses across different computing environments. Future research must establish unified metrics for measuring attack success, common benchmarks for comparing defense effectiveness, and comprehensive evaluation protocols that consider both security and utility aspects. These standardized frameworks should account for practical deployment considerations including resource overhead, performance impact, and scalability of protection methods in real-world scenarios. Additionally, integration challenges across different computing environments must be addressed, as models increasingly operate across multiple deployment contexts simultaneously. This standardization effort requires collaboration between academia and industry to develop meaningful benchmarks that reflect real-world constraints and requirements.

\section{Conclusion} \label{conclusion}
This survey presents the development of model extraction attacks, which attempt to replicate a target model through queries or other observable channels, and organizes them by information channel and computing setting. We show that deployment context shapes both exposure and defense: cloud services must control access while preserving utility, edge devices face physical capture and limited resources, and federated learning, which trains models on distributed clients without centralizing raw data, must protect privacy while enabling collaboration. By linking attack techniques to these environments, our taxonomy clarifies how vulnerabilities arise and which protections are feasible under real constraints. Our assessment of defenses identifies when protection is effective and what costs limit adoption. Case studies show concrete risks in safety critical systems. These findings provide a structured reference on current threats and countermeasures and a basis for adaptive protection that balances security with performance across diverse computing settings.

\bibliographystyle{ACM-Reference-Format}
\bibliography{reference}

\end{document}